\documentclass[aip,pop,reprint,a4paper]{revtex4-1}

\usepackage{amsmath}
\usepackage{amssymb}
\usepackage{accents}
\usepackage{graphicx}
\usepackage{dcolumn}
\usepackage{bm}

\draft 

\begin{document}


\title{The intial value problem in Lagrangian drift kinetic theory} 



\author{J. W. Burby}
 \affiliation{Courant Institute of Mathematical Sciences, New York, New York 10012, USA}


\date{\today}

\begin{abstract}
Existing high-order variational drift kinetic theories contain unphysical rapidly varying modes that are not seen at low-orders. These unphysical modes, which may be rapidly oscillating, damped, or growing, are ushered in by a failure of conventional high-order drift kinetic theory to preserve the structure of its parent model's initial value problem (Vlasov-Poisson for electrostatics, Vlasov-Darwin or Vlasov-Maxwell for electromagnetics.) In short, the system phase space is unphysically enlarged in conventional high-order variational drift kinetic theory. I present an alternative, ``renormalized" variational approach to drift kinetic theory that manifestly respects the parent model's initial value problem. The basic philosophy underlying this alternate approach is that high-order drift kinetic theory ought to be derived by truncating the all-orders system phase space Lagrangian instead of the usual ``field+particle" Lagrangian. For the sake of clarity, this story is told first through the lens of a finite-dimensional toy model of high-order variational drift kinetics; the analogous full-on drift kinetic results are discussed subsequently. The renormalized drift kinetic system, while just as accurate as conventional formulations, does not support the troublesome rapidly varying mode.
\end{abstract}

\pacs{}

\maketitle 

\section{Introduction}
The Lagrangian approach to deriving drift kinetic and gyrokinetic theories\cite{Sugama_2000,Brizard_POP_2000,Brizard_2007} has proven to be successful at producing strongly-magnetized kinetic plasma models with exact energy and momentum conservation laws\cite{Scott_2010}. The philosophy underlying this approach requires all approximations to be implemented at the level of a system Lagrangian. As a result, exact conservation laws are unavoidable, and expressions for the conserved quantities follow systematically from the Noether method.

Conservation laws are not everything, however. Another fundamental property of a physical model is its phase space\cite{Lee_1990}. Phase space, $P$, is defined to be any space in which the model is equivalent to a dynamical law for points $p\in P$ of the form $dp/dt=V(p,t)$, where $V$ is a time-dependent vector field on $P$. Before any mention can be made of a dynamical system's conservation laws, the ``arena" where the dynamics take place must be identified; phase space is precisely this arena.  So while it is natural to demand that kinetic plasma models have correct conservation laws, it is equally natural to require that these models have correct phase spaces. 

The question I will investigate in this paper concerns the fate of phase space in Lagrangian drift kinetic and gyrokinetic theories. While the Lagrangian method gets the conservation laws right, does it produce a model with the correct phase space? Equivalently, do these theories require too many, too few, or just the right number of initial conditions? 

Perhaps surprisingly, the answer to this question depends on the order of accuracy (in $\epsilon=\rho/L$) of the gyrokinetic Lagrangian. At very low orders, the question is not interesting; the answer is ``just the right number." In high-flow drift kinetics, ``very low" means  retaining terms in the guiding center Hamiltonian that are zero'th order in magnetic non-uniformity. In low-flow drift kinetics and gyrokinetics, ``very low" means retaining terms in the Hamiltonian that are second order in magnetic non-uniformity. On the other hand, at higher orders than these specified, the question is much more interesting. In particular, high-flow drift kinetics with a first-order (in magnetic nonuniformity) guiding center Hamiltonian requires special attention.

In Section\,\ref{msfail}, I will show that Ref.\,\onlinecite{Miyato_2009}'s variational high-flow first-order electrostatic drift kinetics, which I will refer to as MSST drift kinetics, has an unphysically large phase space. Where one should only have to specify $F$ at $t=0$ to determine the subsequent evolution of the drift kinetic system, this theory requires $\varphi$ to be supplied independently at $t=0$ as well. As a consequence, the drift kinetic system exhibits certain malignant features. I argue using linearized equations that MSST drift kinetics supports unphysical rapidly-varying modes, akin to the ``runaway" solutions allowed by the Lorentz-Abraham-Dirac equation describing radiation reaction. These modes are rapidly varying in the sense that their frequency increases as $\epsilon$ decreases. Moreover, their presence is entirely a consequence of requiring too many initial conditions; in lower order drift kinetics, where the size of phase space is correct, the modes do not appear.

Fortunately, there is a way to rescue the Lagrangian drift kinetic method from these doldrums. In Section\,\ref{renormalize}, I present a systematic ``renormalized" Lagrangian approach to MSST drift kinetics that manifestly retains the physical phase space. The basic idea underlying this approach is that the system phase space Lagrangian for all-orders drift kinetics is a better starting point for performing asymptotics than the ``particle phase-space Lagrangian $+$ electromagnetic Lagrangian" approach that has become standard among practitioners of Lagrangian gyrokinetic theories. 

Because the derivation of the renormalized Lagrangian involves some relatively advanced techniques from the calculus of variations, I have devoted Section\,\ref{easy} to a description of the necessary manipulations in the context of a finite-dimensional toy model of drift kinetics. This toy model illustrates why standard Lagrangian drift kinetics does violence to the drift kinetic phase space without introducing the infinite-dimensional machinery used in drift kinetic field theory.

It is hard to mention higer-order gyrokinetics without inviting a discussion of the recent momentum conservation controversy (see Ref.\,\onlinecite{Krommes_Hammett_2013} for a detailed description of what the precise disagreements have been.) Without taking a side in the debate, it is fair to say that work related to gyrokinetic momentum conservation has rendered the study of higher-order drift- and gyrokinetic theories much more interesting and important. Because these higher-order theories require an unphysical number of initial conditions when formulated using the standard Lagrangian approach, I believe that the results and techniques outlined in this paper will be important for practitioners of Lagrangian drift- and gyrokinetic theory to understand and apply as the field moves forward. In particular, it will be interesting to derive the appropriate renormalized Lagrangian for various Lagrangian gyrokinetic models, and drift kinetic models that incorporate electromagnetic effects.

\section{Two-oscillator drift kinetic theory\label{easy}}
\subsection{Introducing the all-orders theory}
In this Section I will analyze the initial value problem for a toy model I will call two-oscillator drift kinetics. The motivation for studying this relatively simple model is that its initial value problem is structurally very similar to that of drift kinetics. In this first subsection I will give a derivation of the all-orders version of the model, which starts from the following system of two coupled one-dimensional oscillators,
\begin{align}
\ddot{q}&=-\omega_1^2\,q-2g\,q\varphi\label{pre_TOD_particles}\\
\ddot{\varphi}&=-\omega_2^2\varphi-g q^2.\label{pre_TOD_field}
\end{align}
I will refer to the dynamical system defined by Eqs.\,(\ref{pre_TOD_particles}) and\,(\ref{pre_TOD_field}) as two-oscillator kinetics. The quantities $q,\varphi\in\mathbb{R}$ are the oscillator positions. As the notation suggests, $q$-space should be thought of as an analogue of the single-particle configuration space. Similarly, $\varphi$ should be thought of as the electromagnetic field. The parameters $g,\omega_1,\omega_2$ represent the coupling constant, the characteristic frequency for particles, and the characteristic frequency for the fields, respectively. This system can be derived from the Lagrangian
\begin{align}\label{basic_lag}
&L(q,p,\varphi,\dot{q},\dot{p},\dot{\varphi})=\nonumber\\
L_p(q,&p,\dot{q},\dot{p})+L_{\text{int}}(q,\varphi)+L_f(\varphi,\dot{\varphi}),
\end{align}
where
\begin{align}
L_p&=p\,\dot{q}-\left(\frac{1}{2}p^2+\frac{1}{2}\omega_1^2q^2\right)\\
L_f&=\frac{1}{2}\dot{\varphi}^2-\frac{1}{2}\omega_2^2\varphi^2\\
L_{\text{int}}&=-gq^2\varphi,
\end{align}
and $p$ denotes the particle canonical momentum. Note that the free particle Lagrangian $L_p$ is a so-called ``phase-space Lagrangian." In place of $L_p$, the ``configuration-space Lagrangian," $\tilde{L}_p=\dot{q}^2/2-\omega_1^2q^2/2$, could have been substituted, but the phase space Lagrangian allows greater freedom in applying coordinate transformations to the particle variables. It is for this same reason that derivations of drift kinetics necessarily employ a phase space Lagrangian for particles. The equations of motion corresponding to this system Lagrangian are found by varying the action $S=\int_{t_1}^{t_2} L\,dt$ while holding the values of $q,p,$ and $\varphi$ fixed at $t_1$ and $t_2$. In particular, extremizing the action leads to the Euler-Lagrange equations,
\begin{align}
\frac{\partial L}{\partial q}-\frac{d}{dt}\frac{\partial L}{\partial \dot{q}}&=0\\
\frac{\partial L}{\partial p}-\frac{d}{dt}\frac{\partial L}{\partial \dot{p}}&=0\\
\frac{\partial L}{\partial \varphi}-\frac{d}{dt}\frac{\partial L}{\partial \dot{\varphi}}&=0,
\end{align}
which are readily seen to reproduce Eqs.\,(\ref{pre_TOD_particles}) and\,(\ref{pre_TOD_field}).

All-orders two-oscillator drift kinetics is obtained from this ``particle-space" model by passing from particle coordinates $(q,p)$ to ``guiding center coordinates" $(Q,P)$. Just as in drift kinetic theory, the guiding center coordinates are related to the particle coordinates by a transformation of the particle phase space that depends parametrically on the first $N$ time derivatives of the field $\varphi$, where $N$ is some positive integer. Let $j\varphi=(\varphi,\partial_t\varphi,\dots,\partial_t^N\varphi)$ and $T_{j\varphi}:\mathbb{R}^2\rightarrow\mathbb{R}^2$ be the gyrocenter coordinate transformation. By definition,
\begin{align}
(Q,P)=T_{j\varphi}(q,p)\equiv T(q,p;j\varphi).\label{gyro_transform}
\end{align}
Note that this transformation can be thought of as passing into a moving frame on the particle phase space; the frame moves as the field $\varphi$ evolves.

For the sake of concreteness and simplicity, assume $T_{j\varphi}$ is given by
\begin{align}
T_{j\varphi}(q,p)=(q+\epsilon\ddot{\varphi},p),
\end{align}
where $\epsilon$ is a constant with dimensions of inverse frequency squared. The inverse of $T_{j\varphi}$ is therefore
\begin{align}
T_{j\varphi}^{-1}(Q,P)=(Q-\epsilon\ddot{\varphi},P).
\end{align}
Using these explicit expressions for the guiding center transformation, I will now determine the differential equations governing all-orders two-oscillator drift kinetics.

Suppose that $q,p,\varphi$ satisfy the dynamical equations of two-oscillator kinetics. I will find differential equations governing the quantities $Q=q+\epsilon\ddot{\varphi}$, $P=p$, and $\varphi$. Using the field equation for two-oscillator kinetics, it is straightforward to see that $Q$ is related to $q$ and $\varphi$ by
\begin{align}
Q=q-\epsilon g q^2-\epsilon\omega_2^2\varphi,
\end{align} 
which also gives a relationship between the time derivatives of $Q,q$, and $\varphi$ according to
\begin{align}
\dot{Q}=(1-2\epsilon g q)\dot{q}-\epsilon\omega_2^2\dot{\varphi}.
\end{align}
Therefore whenever $q<\frac{1}{2\epsilon g}$,
\begin{align}
q&=\frac{1}{2\epsilon g}\bigg(1-\sqrt{1-4\epsilon g Q-4\epsilon^2\omega_2^2 g\varphi}\bigg)\nonumber\\
&=Q+\epsilon(g Q^2+\omega_2^2\varphi)+O(\epsilon^2)\nonumber\\
&=\hat{q}_{-}(Q,\varphi),
\end{align}
and whenever $q>\frac{1}{2\epsilon g}$,
\begin{align}
q&=\frac{1}{2\epsilon g}\bigg(1+\sqrt{1-4\epsilon g Q-4\epsilon^2\omega_2^2 g\varphi}\bigg)\nonumber\\
&=\frac{1}{\epsilon g}-Q-\epsilon\bigg(g Q^2+\omega_2^2\varphi\bigg)+O(\epsilon^2)\nonumber\\
&=\hat{q}_{+}(Q,\varphi).
\end{align}
Now consider an interval of time when $q<\frac{1}{2\epsilon g}$. Using the previous expressions, $P=p$ and $\dot{q}=p$, it is straightforward to verify that $\dot{Q}$ and $\dot{P}$ are given by
\begin{align}
\dot{P}&=-(\omega_1^2+2 g \varphi)\hat{q}_{-}(Q,\varphi)\label{dotp_atod}\\
\dot{Q}&=(1-2\epsilon g \hat{q}_{-}(Q,\varphi))P-\epsilon\omega_2^2\dot{\varphi}\\
\ddot{\varphi}&=-\omega_2^2\varphi-g\hat{q}^2_{-}(Q,\varphi).\label{dotphi_atod}
\end{align}
This system of equations governs all-orders two-oscillator drift kinetics when $q<\frac{1}{2\epsilon g}$. When $q>\frac{1}{2\epsilon g}$, the governing equations are instead
\begin{align}
\dot{P}&=-(\omega_1^2+2 g \varphi)\hat{q}_{+}(Q,\varphi)\label{Pdot}\\
\dot{Q}&=(1-2\epsilon g \hat{q}_{+}(Q,\varphi))P-\epsilon\omega_2^2\dot{\varphi}\label{Qdot}\\
\ddot{\varphi}&=-\omega_2^2\varphi-g\hat{q}^2_{+}(Q,\varphi)\label{varphiddot}.
\end{align}
Perplexingly, the preceding analysis has identified two sets of evolution equations for all-orders two-oscillator drift kinetics. Equivalently, the analysis produced a multi-valued vector field on $(P,Q,\varphi,\dot{\varphi})$-space. This multi-valuedness does not signal a breakdown of determinism. Instead it indicates that $(P,Q,\varphi,\dot{\varphi})$-space is two-sheeted, which is in stark contrast to the single-sheeted $(p,q,\varphi,\dot{\varphi})$-space. The cut in the sheet occurs along the line in the $(Q,\varphi)$-plane given by
\begin{align}
D=1-4\epsilon g Q-4\epsilon^2\omega_2^2g\varphi=0.
\end{align}
On the ``top" sheet $\hat{q}_{+}$ is used in the equations of motion, while on the ``bottom" sheet $\hat{q}_{-}$ is used. When the system reaches the cut, one of two things happens. If $P=p=\dot{q}>0$, then the subsequent evolution of the system occurs on the top sheet. If $P<0$, then the subsequent evolution of the system occurs on the bottom sheet. With this knowledge of the sheet structure of $(P,Q,\varphi,\dot{\varphi})$-space together with our multi-valued dynamical vector field, the evolution equations for all-orders two-oscillator drift kinetics are completely specified.

I will conclude this subsection by presenting the Lagrangian formulation of the all-orders two-oscillator drift kinetic system. The Lagrangian is found by substituting the guiding center transformation into the two-oscillator kinetics Lagrangian, which gives
\begin{align}\label{aotod_lag}
&\bar{L}(Q,P,\varphi,\dot{Q},\dot{P},\dot{\varphi},\ddot{\varphi},\dddot{\varphi})=\nonumber\\
&L(Q,P,\varphi,\dot{Q},\dot{P},\dot{\varphi})+\epsilon\bigg(  Q\ddot{\varphi}(2g \varphi+\omega_1^2)-P\dddot{\varphi}\bigg)\nonumber\\
&-\frac{\epsilon^2}{2}\ddot{\varphi}^2(2g \varphi+\omega_1^2).
\end{align}
Because this Lagrangian depends on the third time derivative of $\varphi$, its associated Euler-Lagrange equations are given by
\begin{align}
\frac{\partial \bar{L}}{\partial Q}-\frac{d}{dt}\frac{\partial \bar{L}}{\partial \dot{Q}}&=0\\
\frac{\partial \bar{L}}{\partial P}-\frac{d}{dt}\frac{\partial \bar{L}}{\partial \dot{P}}&=0\\
\frac{\partial \bar{L}}{\partial \varphi}-\frac{d}{dt}\frac{\partial \bar{L}}{\partial \dot{\varphi}}+\frac{d^2}{dt^2}\frac{\partial \bar{L}}{\partial\ddot{\varphi}}-\frac{d^3}{dt^3}\frac{\partial\bar{L}}{\partial\dddot{\varphi}}&=0.
\end{align}
It is straightforward to verify that these equations reduce to
\begin{align}
&-\dot{P}-(Q-\epsilon\ddot{\varphi})(2g\varphi+\omega_1^2)=0\label{pdotdot}\\
&-P+\dot{Q}-\epsilon\dddot{\varphi}=0\label{second_equation}\\
&-\ddot{\varphi}-\omega_2^2\varphi-g Q^2&\nonumber\label{third_equation}\\
&+\epsilon\bigg(\dddot{P}+4 g \dot{Q}\dot{\varphi}+4g Q\ddot{\varphi}+\ddot{Q}(2g\varphi+\omega_1^2)\bigg)\nonumber\\
&-\epsilon^2\bigg(\ddddot{\varphi}(2g\varphi+\omega_1^2)+3g\ddot{\varphi}^2+4 g\dot{\varphi}\dddot{\varphi}\bigg)=0.
\end{align}
Using Eq.\,(\ref{pdotdot}), $\dddot{P}$ can be expressed in terms of $Q$, $\varphi$ and time derivatives of $\varphi$. When this expression is substituted into Eq.\,(\ref{third_equation}), all third- and fourth-time derivatives of $\varphi$ exactly cancel, leaving a quadratic equation for $\ddot{\varphi}$. The two roots of this quadratic equation correspond to the two sheets identified earlier, and each expresses $\ddot{\varphi}$ in terms of $Q$ and $\varphi$. When either of the expressions for $\ddot{\varphi}$ are substituted into Eq.\,(\ref{pdotdot}) and Eq.\,(\ref{second_equation}), evolution equations for $P$ and $Q$ emerge. The resulting system of equations is equivalent to the equations just derived by manipulating the two-oscillator kinetic 
equations directly.

\subsection{Initial value problem for the all-orders model}

Because I was able to write the differential equations governing all-orders two-oscillator drift kinetics explicitly, it is not difficult to see that the phase space for all-orders two-oscillator drift kinetics has the same dimension as the underlying particle-space theory, two-oscillator kinetics. In each case, four numbers must be specified at $t=0$ in order to uniquely determine a solution. Interestingly, in passing from particle-space to guiding center phase space, the phase space for the field-particle system becomes folded, i.e. multi-sheeted.

While it is not simple to gain a complete understanding of the dynamical behavior predicted by all-orders two-oscillator drift kinetics, it is straightforward to determine the characteristic frequencies of oscillations about the bottom-sheet equilibrium $(Q,P,\varphi,\dot{\varphi})=0$. Linearizing Eqs.\,(\ref{dotp_atod})--(\ref{dotphi_atod}) about this equilibrium gives
\begin{align}
\frac{d}{dt}\delta P&=-\omega_1^2(\delta Q+\epsilon \omega_2^2\delta\varphi)\\
\frac{d}{dt}\delta Q&=\delta P-\epsilon\omega_2^2\delta\dot{\varphi}\\
\frac{d}{dt}\delta\varphi&=\delta\dot{\varphi}\\
\frac{d}{dt}\delta\dot{\varphi}&=-\omega_2^2\delta\varphi,
\end{align}
which can be written as $\dot{z}=Az$, where $z=(\delta P,\delta Q,\delta \varphi,\delta\dot{\varphi})$ and $A$ is a $4\times4$ matrix. The spectrum of $A$ is given by $\{i\omega_1,-i\omega_1,i\omega_2,-i\omega_2\}$. This indicates that the normal modes of all-orders two-oscillator drift kinetics linearized about the origin are purely oscillatory.

\subsection{Two oscillator drift kinetics and its initial value problem}
Because I did not make any approximations in passing from the underlying particle-space model to the guiding center model, all-orders two-oscillator drift kinetics is an exact theory. The same cannot be said of ordinary drift kinetic theory. In that theory, the coordinate transformation that relates guiding center coordinates to particle coordinates is only known as an asymptotic series in $\epsilon=\rho/L$. As a result, the system Lagrangian for ordinary drift kinetics is not known exactly, but only asymptotically. In particular, the drift kinetic Lagrangian is always expanded in powers of $\epsilon$ and then truncated at some order. 

For the sake of examining the consequences of this incomplete knowledge of the system Lagrangian in drift kinetic theory, let us now expand the two-oscillator drift kinetic Lagrangian in powers of $\epsilon$ and then drop terms that are $O(\epsilon^2)$ or higher, and then re-do the initial value analysis. I will refer to the dynamical system defined by the truncated Lagrangian simply as two-oscillator drift kinetics. The truncated Lagrangian is given by
\begin{align}
&\bar{L}(Q,P,\varphi,\dot{Q},\dot{P},\dot{\varphi},\ddot{\varphi})=\nonumber\\
&L(Q,P,\varphi,\dot{Q},\dot{P},\dot{\varphi})+\epsilon\bigg(  Q\ddot{\varphi}(2g \varphi+\omega_1^2)-P\dddot{\varphi}\bigg),
\end{align}
which again depends on the third time derivative of $\varphi$. The Euler-Lagrange equations are now
\begin{align}
&-\dot{P}-(Q-\epsilon\ddot{\varphi})(2g\varphi+\omega_1^2)=0\label{first_equation_prime}\\
&-P+\dot{Q}-\epsilon\dddot{\varphi}=0\\
&-\ddot{\varphi}-\omega_2^2\varphi-g Q^2&\nonumber\\
&+\epsilon\bigg(\dddot{P}+4 g \dot{Q}\dot{\varphi}+4g Q\ddot{\varphi}+\ddot{Q}(2g\varphi+\omega_1^2)\bigg)=0.\label{third_equation_prime}
\end{align}
As before, Eq.\,(\ref{first_equation_prime}) can be used to derive an expression for $\dddot{P}$ that can be substituted into Eq.\,(\ref{third_equation_prime}) in order to relate $Q$, $\varphi$ and the time derivatives of $\varphi$. In contrast to the all-orders theory, the third- and fourth-derivatives of $\varphi$ do not cancel, and the following evolution equations for two-oscillator drift kinetics emerge:
\begin{align}
\dot{P}&=-(Q	-\epsilon\ddot{\varphi})(2 g\varphi+\omega_1^2)\\
\dot{Q}&=P+\epsilon\dddot{\varphi}\\
\ddddot{\varphi}&=\frac{1}{\epsilon^2(2 g\varphi+\omega_1^2)}\bigg(\ddot{\varphi}+g Q^2+\omega_2^2\varphi-2 \epsilon g Q\ddot{\varphi}\nonumber\\
&~~~~~~~~~~~~~~~~~~~~~~~~~~~~~~~~~~~-2\epsilon^2g(\ddot{\varphi}^2+2\dot{\varphi}\dddot{\varphi})\bigg).
\end{align}
Evidently the phase space for two-oscillator drift kinetic theory is $(P,Q,\varphi,\dot{\varphi},\ddot{\varphi},\dddot{\varphi})$-space. This is clearly in disagreement with the all-orders theory, which has a four-dimensional two-sheeted phase space. The Lagrangian approximation method has failed in a qualitative manner! This failure can also be detected by examining the behavior of two-oscillator drift kinetics linearized about the equilibrium $(P,Q,\varphi,\dot{\varphi},\ddot{\varphi},\dddot{\varphi})=0$. The linearized equations are given by
\begin{align}
\frac{d}{dt}\delta P&=-\omega_1^2(\delta Q-\epsilon \delta\ddot{\varphi})\\
\frac{d}{dt}\delta Q&=\delta P+\epsilon\delta\dddot{\varphi}\\
\frac{d}{dt}\delta\varphi&=\delta\dot{\varphi}\\
\frac{d}{dt}\delta\dot{\varphi}&=\delta\ddot{\varphi}\\
\frac{d}{dt}\delta\ddot{\varphi}&=\delta\dddot{\varphi}\\
\frac{d}{dt}\delta\dddot{\varphi}&=\frac{1}{\epsilon^2\omega_1^2}(\delta\ddot{\varphi}+\omega_2^2\delta\varphi),
\end{align}
which can be written as $\dot{z}=Az$, where $z=(\delta P,\delta Q,\delta \varphi,\delta \dot{\varphi},\delta \ddot{\varphi},\delta \dddot{\varphi})$ and $A$ is a $6\times 6$ matrix. The spectrum of $A$ is given by $\{i\omega_1,-i\omega_1,i\tilde{\omega}_2,-i\tilde{\omega}_2,\gamma_\times,-\gamma_\times\}$, where
\begin{align}
\tilde{\omega}_2&=\omega_2\bigg(1-\frac{1}{2}\omega_1^2\omega_2^2\epsilon^2\bigg)+O(\epsilon^3)\\
\gamma_\times&=\frac{1}{\epsilon\omega_1}\bigg(1+\frac{1}{2}\omega_1^2\omega_2^2\epsilon^2\bigg)+O(\epsilon^3).
\end{align}
In contrast to the all-orders theory, linearized two-oscillator drift kinetics therefore has rapidly-growing normal modes with growth rate $\gamma_\times$.

As I will show in more detail in the next section, conventional Lagrangian drift kinetic theory suffers from problems similar to those that I have highlighted in the context of two-oscillator drift kinetics. In both systems, the root cause of the trouble is the combination of (a) employing a coordinate change on the single-particle phase space that depends on time derivatives of the electromagnetic fields and (b) truncating a ``field+particle" system Lagrangian. 

\subsection{Rescuing phase space}
This article is not meant to paint a depressing picture of Lagrangian drift kinetics. While the flaw with the conventional Lagrangian approach identified in the previous subsection is important to be aware of, it is equally important to know that the flaw is curable. In this subsection I will describe the cure in the context of two-oscillator drift kinetics.

In order to avoid the possibility of destroying the physical structure of phase space without resorting to the all-orders theory, I advocate truncating the phase space Lagrangian for the entire two-oscillator drift kinetic system. See Appendix\,\ref{dressing_why} for a discussion of the precise meaning of the term ``phase space Lagrangian." I will explain why this is preferable to the Lagrangian approach described in the previous subsection by first showing that it works and then giving a brief abstract discussion. 

First I will derive the phase space Lagrangian for all-orders two-oscillator drift kinetics. See Appendix\,\ref{dressing_why}, which is a partial generalization of a discussion found in Ref.\,\onlinecite{Marsden_1998}, for a rigorous justification of the procedure I will use to do this. In the derivation I will restrict attention to the bottom sheet (corresponding to $\hat{q}_{-}$); the top sheet can be treated in the same manner. Consider the free-endpoint variation of the all-orders two-oscillator drift kinetic Lagrangian. A simple calculation shows
\begin{align}
\delta\int_{t_1}^{t_2} \bar{L}\,dt=&\int_{t_1}^{t_2}\bigg(\frac{\partial \bar{L}}{\partial Q}-\frac{d}{dt}\frac{\partial \bar{L}}{\partial \dot{Q}}\bigg)\delta Q\,dt\nonumber\\
+&\int_{t_1}^{t_2}\bigg(\frac{\partial \bar{L}}{\partial P}-\frac{d}{dt}\frac{\partial \bar{L}}{\partial \dot{P}}\bigg)\delta P\,dt\nonumber\\
+&\int_{t_1}^{t_2}\bigg(\frac{\partial \bar{L}}{\partial \varphi}-\frac{d}{dt}\frac{\partial \bar{L}}{\partial \dot{\varphi}}+\frac{d^2}{dt^2}\frac{\partial \bar{L}}{\partial\ddot{\varphi}}-\frac{d^3}{dt^3}\frac{\partial\bar{L}}{\partial\dddot{\varphi}}\bigg)\delta\varphi\,dt\nonumber\\
+&\bigg[\vartheta_Q\delta Q+\vartheta_\varphi\delta\varphi+\vartheta_{\dot{\varphi}}\delta\dot{\varphi}+\vartheta_{\ddot{\varphi}}\delta\ddot{\varphi}\bigg]_{t_1}^{t_2},
\end{align}
where
\begin{align}
\vartheta_Q&=\frac{\partial\bar{L}}{\partial \dot{Q}}\\
\vartheta_\varphi&=\frac{\partial \bar{L}}{\partial \dot{\varphi}}-\frac{d}{dt}\frac{\partial \bar{L}}{\partial \ddot{\varphi}}+\frac{d^2}{dt^2}\frac{\partial \bar{L}}{\partial \dddot{\varphi}}\\
\vartheta_{\dot{\varphi}}&=\frac{\partial \bar{L}}{\partial \ddot{\varphi}}-\frac{d}{dt}\frac{\partial \bar{L}}{\partial \dddot{\varphi}}\\
\vartheta_{\ddot{\varphi}}&=\frac{\partial \bar{L}}{\partial \dddot{\varphi}}.
\end{align}
I will refer to the $1$-form on $(Q,P,\dot{Q},\dot{P},\varphi,\dot{\varphi},\ddot{\varphi},\dddot{\varphi})$-space given by
\begin{align}
\vartheta^\infty=\vartheta_Q\,dQ+\vartheta_\varphi\,d\varphi+\vartheta_{\dot{\varphi}}\,d\dot{\varphi}+\vartheta_{\ddot{\varphi}}\,d\ddot{\varphi}
\end{align}
as the \emph{bare} all-orders Lagrange $1$-form. As described in Ref.\,\onlinecite{Marsden_1998}, if the Lagrangian $\bar{L}$ for all-orders two-oscillator drift kinetics were non-degenerate, then $\vartheta^\infty$ would give the ``symplectic part" of the system phase space Lagrangian on $(Q,P,\dot{Q},\dot{P},\varphi,\dot{\varphi},\ddot{\varphi},\dddot{\varphi})$-space. However, because all higher time derivatives of $\varphi$ than $\ddot{\varphi}$ cancel out of the Euler-Lagrange equations, $\bar{L}$ is degenerate, and I cannot deduce the phase space Lagrangian so easily. Moreover, it is a good thing that this technique does not work to determine the symplectic part of the system phase space Lagrangian because $(Q,P,\dot{Q},\dot{P},\varphi,\dot{\varphi},\ddot{\varphi},\dddot{\varphi})$-space is not the correct phase space for all-orders two-oscillator drift kinetics! The correct symplectic part of the phase space Lagrangian is given by the \emph{dressed} Lagrange $1$-form $\tilde{\vartheta}^\infty$. Here ``dressing" means using the all-orders two-oscillator drift kinetic equations of motion, Eqs.\,(\ref{Pdot})-\,(\ref{varphiddot}), to write the bare Lagrange $1$-form in terms of only the true phase space variables $(Q,P,\varphi,\dot{\varphi})$. The dressed Lagrange $1$-form is given by
\begin{align}
\tilde{\vartheta}^\infty=\frac{P}{\sqrt{D(Q,\varphi)}}\,dQ+\bigg(\dot{\varphi}+\frac{\epsilon\omega_2^2P}{\sqrt{D(Q,\varphi)}}\bigg)\,d\varphi.
\end{align}
The phase space action for all-orders two-oscillator drift kinetics is therefore
\begin{align}\label{TOD_var}
S(\gamma)=\int_\gamma\tilde{\vartheta}^\infty-\int_{t_1}^{t_2}\tilde{H}^\infty(\gamma(t))\,dt,
\end{align}
where $\gamma$ is curve in $(Q,P,\varphi,\dot{\varphi})$-space parameterized by the time interval $[t_1,t_2]$ and $\tilde{H}^\infty$ is the dressed Hamiltonian. The dressed Hamiltonian is obtained from the usual ``bare" Hamiltonian,
\begin{align}
&H(Q,P,\dot{Q},\dot{P},\varphi,\dot{\varphi},\ddot{\varphi},\dddot{\varphi})=\frac{\partial\bar{L}}{\partial \dot{Q}}\dot{Q}+\frac{\partial\bar{L}}{\partial \dot{P}}\dot{P}+\frac{\partial\bar{L}}{\partial \dddot{\varphi}}\dddot{\varphi}\nonumber\\
&+\bigg(\frac{\partial\bar{L}}{\partial \ddot{\varphi}}-\frac{d}{dt}\frac{\partial\bar{L}}{\partial \dddot{\varphi}}\bigg)\ddot{\varphi}+\bigg(\frac{\partial\bar{L}}{\partial \dot{\varphi}}-\frac{d}{dt}\frac{\partial\bar{L}}{\partial \ddot{\varphi}}+\frac{d^2}{dt^2}\frac{\partial\bar{L}}{\partial \dot{\varphi}}\bigg)\dot{\varphi}\nonumber\\
&-\bar{L},
\end{align}
by using the all-orders two-oscillator drift kinetic equations of motion, Eqs.\,(\ref{Pdot})-\,(\ref{varphiddot}) to express the bare Hamiltonian in terms of the true phase space variables $(Q,P,\varphi,\dot{\varphi})$. A straightforward calculation shows that the dressed Hamiltonian is given by
\begin{align}\label{TOD_dham}
\tilde{H}^\infty(Q,P,\varphi,\dot{\varphi})&=\frac{1}{2}\dot{\varphi}^2+\frac{1}{2}P^2+\frac{1}{2}\omega_2^2\varphi^2\nonumber\\
&+\frac{1}{2}\hat{q}_{-}^2(Q,\varphi)(\omega_1^2+2g\varphi).
\end{align}

By applying fixed-endpoint variations to the phase space action, it is straightforward to verify directly that all-orders two-oscillator drift kinetics admits this alternative variational principle. Indeed, for a general dressed Hamiltonian the Euler-Lagrange equations give
\begin{align}
\frac{d}{dt}Q&=\sqrt{D}\frac{\partial \tilde{H}^\infty}{\partial P}-\epsilon\omega_2^2\frac{\partial \tilde{H}^\infty}{\partial \dot{\varphi}}\\
\frac{d}{dt}P&=-\sqrt{D}\frac{\partial \tilde{H}^\infty}{\partial Q}\\
\frac{d}{dt}\varphi&=\frac{\partial\tilde{H}^\infty}{\partial \dot{\varphi}}\\
\frac{d}{dt}\dot{\varphi}&=-\frac{\partial\tilde{H}^\infty}{\partial \varphi}+\epsilon\omega_2^2\frac{\partial\tilde{H}^\infty}{\partial Q}.
\end{align}
When the physical dressed Hamiltonian given in Eq.\,(\ref{TOD_dham}) is substituted into this last equation set, the dynamical equations defining all-orders two-oscillator drift kinetics (on the bottom sheet) are recovered.

With this phase space action principle for all-orders two-oscillator drift kinetics in hand, I will now use it to construct an alternative truncated theory that I will refer to as \emph{renormalized} two-oscillator drift kinetics (I will explain the use of the term ``renormalized" shortly.) Renormalized two-oscillator drift kinetics is defined simply by replacing $\tilde{\vartheta}^\infty$ and $\tilde{H}^\infty$ in Eq.\,(\ref{TOD_var}) with the corresponding first-order Taylor expansions in $\epsilon$, $\tilde{\vartheta}$ and $\tilde{H}$, where
\begin{align}
\tilde{\vartheta}&=P(1+\epsilon 2 gQ)\,dQ+(\dot{\varphi}+\epsilon P\omega_2^2)\,d\varphi\\
\tilde{H}&=\frac{1}{2}\dot{\varphi}^2+\frac{1}{2}P^2+\frac{1}{2}\omega_2^2\varphi^2+\frac{1}{2}(\omega_1^2+2g\varphi)Q^2\nonumber\\
&~~+\epsilon(\omega_1^2\omega_2^2 Q\varphi+2g\omega_2^2 Q\varphi^2+[2g^2\varphi+g\omega_1^2]Q^3).
\end{align}
Thus the phase space action for renormalized two-oscillator drift kinetics is given by
\begin{align}\label{rTOD_var}
S(\gamma)=\int_\gamma\tilde{\vartheta}-\int_{t_1}^{t_2}\tilde{H}(\gamma(t))\,dt,
\end{align}
and the associated equations of motion are
\begin{align}
\frac{d}{dt}Q&=\frac{1}{1+2\epsilon g Q}\bigg(\frac{\partial \tilde{H}}{\partial P}-\epsilon\omega_2^2\frac{\partial \tilde{H}}{\partial \dot{\varphi}}\bigg)\\
\frac{d}{dt}P&=-\frac{1}{1+2\epsilon g Q}\frac{\partial \tilde{H}}{\partial Q}\\
\frac{d}{dt}\varphi&=\frac{\partial\tilde{H}}{\partial \dot{\varphi}}\\
\frac{d}{dt}\dot{\varphi}&=-\frac{\partial\tilde{H}}{\partial \varphi}+\frac{\epsilon\omega_2^2}{1+2\epsilon g Q}\frac{\partial\tilde{H}}{\partial Q}.
\end{align}
It is simple to check that the dynamical vector field for renormalized two-oscillator drift kinetics agrees with the ``bottom sheet" dynamical vector field for all-orders two-oscillator drift kinetics with $O(\epsilon)$ accuracy. Moreover renormalized two-oscillator drift kinetics is superior to two-oscillator drift kinetics because it is defined on the correct phase space. This last point is illustrated clearly by considering the linearization of renormalized two-oscillator drift kinetics about the equilibrium $(Q,P,\varphi,\dot{\varphi})=0$. The linearized equations are given by
\begin{align}
\frac{d}{dt}\delta Q&=\delta P-\epsilon \omega_2^2 \delta\dot{\varphi}\\
\frac{d}{dt}\delta P&=-\omega_1^2(\delta Q+\epsilon\omega_2^2\delta\varphi)\\
\frac{d}{dt}\delta \varphi&=\delta\dot{\varphi}\\
\frac{d}{dt}\delta\dot{\varphi}&=-\omega_2^2(1-\epsilon^2\omega_1^2\omega_2^2)\delta\varphi,
\end{align}
which can be written as $\dot{z}=Az$, where $z=(\delta Q,\delta P,\delta\varphi,\delta\dot{\varphi})$ and $A$ is a $4\times 4$ matrix. The spectrum of $A$ is given by$\{i\omega_1,-i\omega_1,i\omega_2\sqrt{1-\epsilon^2\omega_1^2\omega_2^2},-i\omega_2\sqrt{1-\epsilon^2\omega_1^2\omega_2^2}\}$, which implies that all normal modes of linearized renormalized two-oscillator drift kinetics are oscillatory, just as in the all-orders theory. Moreover, the normal mode frequencies of the renormalized theory are close to those of the all-orders theory.

How did did this modified Lagrangian procedure succeed in preserving the correct phase space? The answer is that by implementing approximations at the level of the system phase space Lagrangian it is impossible for the phase space of all-orders two-oscillator drift kinetics to be modified (modulo non-perturbative modifications of the phase space sheet structure.) As long as the $1$-form appearing in the phase space action has a non-degenerate exterior derivative, the Euler-Lagrange equations that follow from the phase space action principle are guaranteed to define a first-order ODE on the space where the $1$-form is defined. This non-degeneracy condition is morally guaranteed by the fact that the transformation between the particle coordinates and the guiding center coordinates is near-identity.

In what sense is renormalized two-oscillator drift kinetics a ``renormalized" theory? While I do not believe the renormalization group is playing any role here, the procedure described in this subsection bears a strong resemblance to the procedure used to extract sensible results from perturbative quantum field theories. I began with a system with a nonsensical phase space, namely two-oscillator drift kinetics. In a loose sense, this system contained an unphysical infinity. To eliminate the absurd features of this model, I regularized it by passing to all-orders two-oscillator drift kinetics. This step can be thought of as inserting a ``cutoff"; I inserted higher-order $\epsilon$-dependent terms into the two-oscillator drift kinetic Lagrangian that guaranteed the dimension of phase space was sensible. Finally, I devised a method to vary the cutoff (i.e. change which higher-order terms were in the theory) that would not alter the value of the dimension of phase space. This step is analogous to tailoring the values of the coupling constants in a perturbative quantum field theory\cite{Wentzel_QFT_1949} in order to render the propagator cutoff-independent.

\section{Conventional Lagrangian electrostatic drift kinetics and its flaws\label{msfail}}
In passing from two-oscillator kinetics to two-oscillator drift kinetics, the size of phase space was increased for unphysical reasons. In this Section I will argue that the same phenomenon occurs in a variant of first-order high-flow electrostatic drift kinetics that I will refer to as MSST drift kinetics (the acronym corresponds to the names of the authors of Ref.\,\onlinecite{Miyato_2009}.) MSST drift kinetics is defined by the system Lagrangian 
\begin{align}
&L(F,V,\varphi,\dot{\varphi})=\nonumber\\
&\sum_s\int \ell_s(z,V_s(z);\varphi,\dot{\varphi})\,F_s\,d\Omega_s\nonumber\\
&+\frac{\epsilon_o}{2}\int|\nabla\varphi|^2\,d^3\bm{X},
\end{align}
where the guiding center phase space Lagrangian $\ell_s$ is the one given in Eq.\,(45) of Ref.\,\onlinecite{Miyato_2009}, 
\begin{align}
\ell_s(z,\dot{z};\varphi,\dot{\varphi})=&\left(q_s\bm{A}+m_s v_\parallel \hat{b}-\frac{m_s}{q_s}\mu \bm{W}\right)\cdot\dot{\bm{X}}\nonumber\\
&+\frac{q_s}{m_s}\mu\,\dot{\zeta}-h_s(z;\varphi,\bm{E},\dot{\bm{E}}),
\end{align}
with
\begin{align}\label{MSSTH}
h_s(z;\varphi,\bm{E},\dot{\bm{E}})=q_s\varphi+K_s(\bm{E},\dot{\bm{E}}),
\end{align}
and $K_s(\bm{E},\dot{\bm{E}})=K_{os}(\bm{E})+K_{1s}(\bm{E},\dot{\bm{E}})$ is given by
\begin{align}
K_{os}(\bm{E})&=\frac{1}{2}m_sv_\parallel^2+\mu B-\frac{1}{2}m_su_E^2\label{Ka}\\
K_{1s}(\bm{E},\dot{\bm{E}})&=\frac{1}{2}\frac{m_s}{q_s}\bigg(\mu+\frac{m_su_E^2}{2B}\bigg)\hat{b}\cdot\nabla\times\bm{u}_E\nonumber\\
&-\frac{m_s v_\parallel}{2\omega_{cs}}\bm{u}_E\cdot\nabla\times\bm{u}_E\nonumber\\
&+\frac{7}{6}\frac{m_s}{q_s}\mu(\hat{b}\times\bm{u}_E)\cdot\nabla\text{ln}\,B\nonumber\\
&+\frac{m_s u_E^2 v_\parallel}{\omega_{cs}}\tau+\frac{m_s v_\parallel^2}{\omega_{cs}}\hat{b}\times\bm{u}_E\cdot\bm{\kappa}\nonumber\\
&-\frac{q_s}{2\omega_{cs}^2}\bm{u}_E\cdot\dot{\bm{E}}\label{Kb}.
\end{align}
Here $F$ is the (multi-species) distribution function; $V$ is the Eulerian phase space fluid velocity; $\varphi$ is the dynamical electrostatic potential; $z=(\bm{X},v_\parallel,\mu,\zeta)$ is a point in the guiding center phase space consisting of the guiding center position, parallel velocity, magnetic moment, and gyrophase; $\bm{A}$ is the background vector potential; $\hat{b}$ is the unit vector along the background magnetic field; $\tau=\hat{b}\cdot\nabla\times\hat{b}$ and $\bm{\kappa}=\hat{b}\cdot\nabla\hat{b}$; $\bm{W}=(\nabla \hat{e}_1)\cdot \hat{e}_2+\tau\hat{b}/2$, where $\hat{e}_1,\hat{e}_2$ are unit vectors perpendicular to the background magnetic field; $\bm{E}=-\nabla\varphi$ is the electric field; $\bm{u}_E=\bm{E}\times\hat{b}/B$ is the $E\times B$ velocity; the Liouville volume element $d\Omega_s=\mathcal{J}_s\,d^3\bm{X}\,dv_\parallel\,d\mu\,d\zeta$; and $\mathcal{J}_s= B_{\parallel s}^*/m_s$, where $\bm{B}_s^*=\nabla\times(\bm{A}+m_s v_\parallel\hat{b}/q_s-m_s\mu\bm{W}/q_s^2)$. The authors of Ref.\,\onlinecite{Miyato_2009} drop a number of terms from $K_{1s}$ by invoking some often-relevant subsidiary orderings (relative to the usual $\epsilon=\rho/L$ ordering.) Here, I will not invoke these subsidiary orderings, and will therefore keep all of the terms in $K_{1s}$.

The equations of motion corresponding to this system Lagrangian may be obtained by varying $\varphi$, $F$, and $V$ in the action integral $S=\int_{t_1}^{t_2}L\,dt$ subject to the constraints
\begin{align}
\delta V_s&=\dot{\xi}_s+L_{V_s}\xi_s\\
\delta F_s&=-\frac{1}{\mathcal{J}_s}\text{div}(\mathcal{J}_s\xi_sF_s),
\end{align}
where $\xi$ is an arbitrary time-dependent (multi-species) vector field on the guiding center phase space and $L_{V_s}$ denotes the Lie derivative along the vector field $V_s$. Varying $\varphi$ gives the Poisson equation
\begin{align}\label{quasineutral_a}
\nabla\cdot\bigg(\epsilon_o\bm{E}-\frac{\delta\mathcal{K}}{\delta\bm{E}}+\partial_t\frac{\delta\mathcal{K}}{\delta\dot{\bm{E}}}\bigg)=\sum_{s}q_s N_s
\end{align}
where $\mathcal{K}=\mathcal{K}(F,\bm{E},\dot{\bm{E}})$ is given by
\begin{align}
\mathcal{K}(F,\bm{E},\dot{\bm{E}})=\sum_s\int K_s(\bm{E},\dot{\bm{E}})F_s\,d\Omega_s,
\end{align} 
\begin{align}
N_s=\int F_s\,\mathcal{J}_s\,dv_\parallel\,d\mu\,d\zeta,
\end{align}
and $\omega_{cs}=q_s B/m_s$. Varying $V$ and $F$ gives the guiding center equations of motion
\begin{align}
V_s^i&=\{z^i,H_s\}_s^{\text{gc}},
\end{align}
where $\{\cdot,\cdot\}^{\text{gc}}$ denotes the guiding center Poisson bracket associated with the guiding center phase space Lagrangian $\ell_s$. Explicitly, 
\begin{align}
\dot{\bm{X}}_s&=\frac{\bm{B}_s^*}{m_s B_{\parallel s}^*}\frac{\partial H_s}{\partial v_\parallel}+\frac{\hat{b}\times\nabla H_s}{q_s B_{\parallel s}^*}\label{dotX}\\
\dot{v}_{\parallel s}&=-\frac{\bm{B}_s^*\cdot\nabla H_s}{m_s B_{\parallel s}^*}.
\end{align}
The distribution function $F$ therefore satisfies
\begin{align}\label{vlasov_a}
\frac{\partial F_s}{\partial t}+\frac{1}{\mathcal{J}_s}\nabla\cdot(\dot{\bm{X}}_sF_s\mathcal{J}_s)+\frac{1}{\mathcal{J}_s}\frac{\partial}{\partial v_\parallel}(\dot{v}_{\parallel s}F_s\mathcal{J}_s)=0.
\end{align}

The dynamical system specified by Eqs.\,(\ref{quasineutral_a}) and (\ref{vlasov_a}) is meant to approximate the Vlasov-Poisson system in a strong background magnetic field when the time and length scales of the electrostatic potential are long compared with the gyrofrequency and gyroradius, respectively. In particular, because the electrostatic potential in the Vlasov-Poisson system is a functional of the distribution function, Eqs.\,(\ref{quasineutral_a}) and (\ref{vlasov_a}) are supposed to approximate a system of equations whose phase space is the space of $F$'s. Moreover, I will argue in the next section that when high-flow electrostatic gyrokinetics is carried to all orders in magnetic nonuniformity, the phase space is indeed the space of $F$'s. It is therefore reasonable to hope that Eqs.\,(\ref{quasineutral_a}) and (\ref{vlasov_a}) are equivalent to a first-order ODE on the space of $F$'s. 

Unforunately this hope is ill-founded. I will explain why in the context of linearized MSST drift kinetics. For simplicity's sake, let the background magnetic field be uniform and $z$-directed. The equilibrium I will linearize about satisfies $\varphi=\dot{\varphi}=0$ and
\begin{align}\label{thermal_eq}
F_{os}=\left(\frac{m_s}{2\pi T}\right)^{3/2}N_{os}(\bm{X})\exp(-(\mu B+m_s v_\parallel^2/2)/T),
\end{align}
where $T$ is the equilibrium temperature and $N_{os}$ is the equilibrium number density. This type of equilibrium will exist if and only if
\begin{gather}
-\nabla\cdot\nabla_\perp\bigg(\sum_sq_s\rho_s^2 N_{os}\bigg)=\sum_sq_s N_{os}\\
\hat{b}\cdot\nabla N_{os}=0,
\end{gather}
where $\rho_s^2\equiv\frac{1}{2}\frac{T/m_s}{\omega_{cs}^2}$ is a typical squared gyroradius of a species-$s$ particle. Linearization of MSST drift kinetics about this type of equilibrium leads to the dynamical equations
\begin{gather}
-\epsilon_o\Delta_\perp\delta\varphi=\sum_s q_s \delta N_s+\nabla_\perp^2\left(\sum_s\frac{\delta M_s}{2\omega_{cs}}\right)\nonumber\\
+\epsilon_o\nabla\tau_o\times\hat{b}\cdot\nabla\delta\dot{\varphi}\label{dkpoisson}\\
\delta\dot{F}_{s}+v_\parallel \hat{b}\cdot \nabla \delta F_s+\frac{\hat{b}}{q_s B}\times\nabla\bigg(q_s\delta\varphi+\frac{\mu \nabla_\perp^2\delta\varphi}{2\omega_{cs}}\bigg)\cdot\nabla F_{so}\nonumber\\
-\frac{\hat{b}}{m_s}\cdot\nabla\bigg(q_s\delta\varphi+\frac{\mu \nabla_\perp^2\delta\varphi}{2\omega_{cs}}\bigg)\frac{\partial F_{os}}{\partial v_\parallel}=0,
\end{gather}
where the elliptic linear operators $\Delta_\perp$ and $\nabla_\perp^2$ are given by
\begin{align}
\Delta_\perp u&=\Delta u+\nabla\cdot\bigg(\sum_s\frac{c^2\mu_o m_s N_{os}}{B^2}\nabla_\perp u\bigg)\\
\nabla_\perp^2 u&=\nabla\cdot\nabla_\perp u,
\end{align}
and the quantities $\delta M_s,\tau_o$ are the perturbed magnetic moment density,
\begin{align}
\delta M_s=\int \mu \delta F_s \frac{B}{m_s}\,dv_\parallel\,d\mu\,d\zeta,
\end{align}
and
\begin{align}
\tau_o=\sum_s\frac{q_s N_{os}}{\epsilon_o\omega_{cs}^2B}.
\end{align}
Instead of analyzing these equations in their full generality, I will restrict attention to proton-electron plasmas with equilibrium density given by
\begin{align}
N_{os}=n\bigg(1+\frac{x}{L}\bigg),
\end{align}
that are uniform along the $z$-axis. In this special case, the linearized MSST equations can be reduced to a system of equations involving only the perturbed electrostatic potential,
\begin{align}\label{doom}
-\epsilon_o\Delta_\perp\delta\dot{\varphi}-(\gamma\nabla_\perp^4+\alpha\nabla_\perp^2)\partial_y\delta\varphi+\beta\partial_y\delta\ddot{\varphi}=0,
\end{align}
where
\begin{align}
\alpha&=\sum_s\frac{nT}{B^2\omega_{cs}L}\\
\beta&=\sum_s\frac{m_s n}{B^2\omega_{cs}L}\\
\gamma&=\sum_s\frac{nT^2}{2 B^2\omega_{cs}^3 L m_s},
\end{align}
\begin{align}
\Delta_\perp\delta\dot{\varphi}=\frac{c^2}{v_A^2}\nabla_\perp^2\delta\dot{\varphi}+\frac{c^2}{v_{Ao}^2L}\partial_x\delta\dot{\varphi},
\end{align}
and 
\begin{align}
\frac{c^2}{v_{Ao}^2}&=\sum_s\frac{c^2\mu_o m_s n}{B^2}\\
\frac{c^2}{v_A^2}&=1+\frac{c^2}{v_{Ao}^2}\bigg(1+\frac{x}{L}\bigg).
\end{align}
A thorough analysis of Eq.\,(\ref{doom}), which is linear with non-constant coefficients, requires specifying boundary conditions on $\delta\varphi$. However, in order to reveal the nature of modes that vary rapidly in space and time, it is enough to find the dispersion relation in the WKB approximation. Set $\delta\varphi=a \exp(i S)$ where $a$ is a slowly varying amplitude and $S$ is a rapidly varying phase function. Define $\omega=\partial_t S$ and $\bm{k}=\nabla S$. Upon substituting this ansatz into Eq.\,(\ref{doom}) and then dropping all terms involving gradients of the amplitude $a$, the following dispersion relation emerges
\begin{gather}
-\beta k_y\omega^2+\epsilon_o\bigg(\frac{c^2}{v_A^2}(k_x^2+k_y^2)+\frac{c^2}{v_{Ao}^2}\frac{k_x}{iL}\bigg)\omega\nonumber\\
+\bigg(\alpha(k_x^2+k_y^2)+\gamma(k_x^2+k_y^2)^2\bigg)k_y=0.
\end{gather} 
Upon noting that $\alpha,\beta=O(\epsilon)$ and that $\gamma=O(\epsilon^3)$, it becomes clear that when $k_x\ll k_y$ the two branches of this dispersion relation are given to leading order in $\epsilon$ as
\begin{align}
\omega_d&\approx-\frac{v_* k_y}{1+\frac{x}{L}+\frac{v_{Ao}^2}{c^2}}\label{drift_wave}\\
\omega_\times&\approx k_y\frac{T/m_i}{v_*}\bigg(1+\frac{x}{L}+\frac{v_{Ao}^2}{c^2}\bigg)\label{bad_wave},
\end{align}
where $v_*=T/(q_i BL)=O(\epsilon)$ is the so-called diamagnetic drift velocity. The first branch, $\omega_d$, reproduces the well-known drift wave dispersion relation. The second branch, $\omega_\times$, is unphysical because its frequency increases as $\epsilon\rightarrow 0$\footnote{This implies that the wave violates the drift kinetic approximation, namely $\omega\ll\omega_c$, where $\omega$ is any characteristic frequency of interest.}. The unphysical branch is ushered into the theory by the appearance of the second time derivative of $\delta\varphi$ in Eq.\,(\ref{doom}), which itself appears as a consequence of the first time derivative of $\delta\varphi$ appearing in the linearized drift kinetic Poisson equation, Eq.\,(\ref{dkpoisson}). In short, the presence of the unphysical mode is a direct consequence of the impossibility of expressing MSST drift kinetics as a first-order ODE in $F$-space. In fact we have seen that MSST drift kinetics reduces to a first-order ODE in $(\varphi,F)$-space when linearized about the class of slab equilirbia just considered, which is larger than $F$-space in the sense that $F$-space is a proper subset of $(\varphi,F)$-space. In the following section I will construct an alternative variational drift kinetic theory that is just as accurate (in terms of $\epsilon$) as MSST drift kinetics, but that does not suffer from an unphysically-large phase space and its associated unphysical rapidly varying modes.

\section{Renormalized MSST drift kinetics\label{renormalize}}
\subsection{The all-orders phase space Lagrangian}
The all-orders version of MSST drift kinetics is derived as follows. The starting point is the magnetized Vlasov-Poisson system expressed in a special coordinate system on single-particle phase space,
\begin{align}
&\partial_t f_s+\nabla\cdot([\bm{u}_E+\bm{w}]f_s)+\nabla_{\bm{w}}\cdot(\dot{\bm{w}}_sf_s)=0\\
&\epsilon_o\Delta\varphi=-\rho(f)\\
&\dot{\bm{w}}_s=-\partial_t\bm{u}_E-\bm{u}_E\cdot\nabla\bm{u}_E-\bm{w}\cdot\nabla\bm{u}_E+\frac{q_s}{m_s}\bm{w}\times\bm{B},
\end{align}
where $\rho(f)=\sum_sq_s\int f_s\,d^3\bm{w}$, $\bm{B}$ is a static nowhere vanishing magnetic field, the velocity-like variable $\bm{w}$ is related to the single-particle velocity $\bm{v}$ by $\bm{v}=\bm{u}_E+\bm{w}$, and $f$ is the distribution function on $(\bm{x},\bm{w})$-space. For simplicity, I will assume the spatial domain is all of $\mathbb{R}^3$ and that we seek square integrable electrostatic potentials $\varphi$ with square integrable gradients. In this space of amissible $\varphi$, the Laplacian $\Delta$ is invertible, and I will denote its inverse $G=\Delta^{-1}$. Using $G$, the magnetized Vlasov-Poisson system can be expressed as the ODE on $f$-space,
\begin{align}
\partial_t f_s&=-\nabla\cdot(\dot{\bm{x}}(f)f_s)-\nabla_{\bm{w}}\cdot(\dot{\bm{w}}_s(f)f_s)\\
&\equiv\dot{f}_s(f),
\end{align}
where
\begin{align}
&\dot{\bm{x}}(f)=\bm{u}_E(f)+\bm{w}\\
&\bm{a}_s(f)=-\dot{\bm{u}}_E(f)-\bm{u}_E(f)\cdot\nabla\bm{u}_E(f)-\bm{w}\cdot\nabla\bm{u}_E(f)\nonumber\\
&\quad\quad\quad+\frac{q_s}{m_s}\bm{w}\times\bm{B},
\end{align}
and the vectors $\bm{u}_E(f)=\bm{\mathcal{E}}(f)\times b/B$ and $\dot{\bm{u}}_E(f)=\bm{\mathcal{E}}^{(1)}(f)\times b/B$ are determined by the operators
\begin{align}
\bm{\mathcal{E}}(f)&=\frac{1}{\epsilon_o}\nabla G[\rho(f)]\\
\bm{\mathcal{E}}^{(1)}(f)&=-\frac{1}{\epsilon_o}\nabla G[\nabla\cdot(\bm{u}_E(f)\,\rho(f)+\bm{J}(f))]
\end{align}
with $\bm{J}(f)=\sum_sq_s\int \bm{w} f_s\,d^3\bm{w}$. Note that the magnetized Vlasov-Poisson equations imply $\partial_t\rho(f)=-\nabla\cdot(\bm{u}_E(f)\rho(f)+\bm{J}(f))$.

Next, a near-identity field-dependent phase-space coordinate transformation, $\Phi_s(j\bm{E}):P\rightarrow P$, is identified that decouples the rapid gyromotion from the slower drift motion. Here $j\bm{E}=(\bm{E},\partial_t\bm{E},\partial_t^2\bm{E},\dots)$ is an infinite sequence of vector fields, and $P$ is the $6$-dimensional position-velocity phase space. There are many possible ways to choose this coordinate transformation, and MSST drift kinetics is defined by the particular choice described in Ref.\,\onlinecite{Miyato_2009}. 

Using this coordinate transformation, the guiding center distribution function $F$ is defined as
\begin{align}\label{gc_dist}
F_s=\Phi_{s}(j\bm{E})_*f_s,
\end{align}
where $f_s$ is the particle distribution function, and the subscripted $*$ denotes the pushforward. Assuming that $\bm{E}$ and $f_s$ satisfy the magnetized Vlasov-Poisson equations, the right-hand-side of Eq.\,(\ref{gc_dist}) can be expressed entirely in terms of $f$ because $j\bm{E}=(j\bm{\mathcal{E}})(f)=(\bm{\mathcal{E}}(f),\bm{\mathcal{E}}^{(1)}(f),\bm{\mathcal{E}}^{(2)}(f),\dots)$, where 
\begin{align}
\bm{\mathcal{E}}(f)&=\frac{1}{\epsilon_o}\nabla G[\rho(f)]\\
\bm{\mathcal{E}}^{(1)}(f)&=-\frac{1}{\epsilon_o}\nabla G[\nabla\cdot(\bm{u}_E(f)\,\rho(f)+\bm{J}(f))]\\
\bm{\mathcal{E}}^{(2)}(f)&=\dots,
\end{align}
and the operators $\mathcal{E}^{(n)}$ for $n>1$ are determined by expressing the $n$'th time derivative of $\bm{E}$ in terms of $f$ using the magnetized Vlasov-Poisson equations. To be precise,
\begin{align}
F_s&=\Phi_s(j\bm{\mathcal{E}}(f))_*f_s\\
&\equiv\mathcal{F}_s^{-1}(f).
\end{align}
Note that $\mathcal{F}^{-1}$ maps multi-species functions on $P$ into multi-species functions on $P$; formally $\mathcal{F}^{-1}:\oplus_s C^{\infty}(P)\rightarrow\oplus_s C^{\infty}(P)$, where $\oplus_s$ denotes the species-sum of vector spaces.

Because $\Phi$ is near-identity, $\mathcal{F}_s^{-1}(f)\approx f_s$. This implies that $\mathcal{F}^{-1}$ has a unique perturbatively-defined inverse $\mathcal{F}$ that satisfies
\begin{align}\label{calF}
F_s=\Phi_s(j\bm{\mathcal{E}}(\mathcal{F}(F)))_*\mathcal{F}_s(F).
\end{align}
I will explain the procedure for obtaining an asymptotic expansion of $\mathcal{F}$ in the next subsection. Therefore,
\begin{align}
&\partial_t\mathcal{F}(F)=\partial_tf\\
\Rightarrow&D\mathcal{F}[\partial_t F]=\dot{f}(f)\\
\Rightarrow&\partial_t F=D\mathcal{F}^{-1}[\dot{f}(\mathcal{F}(F))].\label{all_orders_eqn}
\end{align}
Here $D\mathcal{F}$ denotes the Fr\'echet derivative of $\mathcal{F}$. Assuming that $\Phi$ is chosen according to the prescription given in Ref.\,\onlinecite{Miyato_2009}, Eq.\,(\ref{all_orders_eqn}) is the all-orders version of MSST drift kinetics. Because this system is manifestly an ODE on $F$-space, the phase space for all-orders MSST drift kinetics is $F$-space, which proves my earlier claim. 

A variational principle for all-orders MSST drift kinetics can be derived by transforming the following variational principle for the magnetized Vlasov-Poisson system into guiding center coordinates. The Vlasov-Poisson Lagrangian is given by
\begin{align}
\mathfrak{L}(f,U,\varphi,\bm{E})=&\sum_s\int l_s(z,U_s(z),\varphi,\bm{E})\,f_s\,d\omega\\\
&+\frac{\epsilon_o}{2}\int|\nabla\varphi|^2\,d^3\bm{x},
\end{align}
where the single-particle phase space Lagrangian is
\begin{align}
l_s(z,\dot{z},\varphi,\bm{E})=&(m_s\bm{w}+m_s\bm{u}_E+q_s\bm{A})\cdot\dot{\bm{x}}\nonumber\\
&-H_s(z,\varphi,\bm{E}),
\end{align}
the single-particle Hamiltonian is
\begin{align}
H_s(z,\varphi,\bm{E})=\frac{1}{2}m_s (\bm{w}+\bm{u}_E)^2+q_s\varphi,
\end{align}
the volume element $d\omega=d^3\bm{x}\,d^3\bm{w}$, and $U$ is the Eulerian phase space fluid velocity. The equations of motion corresponding to this system Lagrangian may be obtained by varying $\varphi$, $f$, and $U$ in the action integral $S=\int_{t_1}^{t_2}L\,dt$ subject to the constraints
\begin{align}
\delta U_s&=\dot{\xi}_s+L_{U_s}\xi_s\\
\delta f_s&=-\text{div}(\xi_sf_s),
\end{align}
where $\xi$ is an arbitrary time-dependent vector field on the (multi-species) guiding center phase space and $L_{U_s}$ denotes the Lie derivative along the vector field $U_s$.

Using the MSST prescription for choosing $\Phi$, the transformed system Lagrangian is given by
\begin{align}\label{bad_lag}
L^{\infty}(F,V,\varphi)=&\sum_s\int \ell^{\infty}_s(z,V_s(z),\varphi,j\bm{E})\,F_s\,d\Omega_s\nonumber\\
&+\frac{\epsilon_o}{2}\int|\nabla\varphi|^2\,d^3\bm{x},
\end{align}
where the all-orders guiding center phase space Lagrangian is given by
\begin{align}
\ell^\infty_s(z,\dot{z},\varphi,j\bm{E})=&\left(q_s\bm{A}+m_s v_\parallel \hat{b}-\frac{m_s}{q_s}\mu \bm{W}\right)\cdot\dot{\bm{X}}\nonumber\\
&+\frac{q_s}{m_s}\mu\,\dot{\zeta}-h^\infty_s(z,\varphi,j\bm{E}),
\end{align}
and the all-orders guiding center Hamiltonian is of the form
\begin{align}
h^\infty_s(z,\varphi,j\bm{E})=K_s^\infty(j\bm{E})+q_s\varphi,
\end{align}
where $K_s^\infty=K_{os}+K_{1s}+K_{2s}+\dots$ and $K_{os},K_{1s}$ are given in Eqs.\,(\ref{Ka}) and (\ref{Kb}).
Note that the MSST method for finding the guiding center transformation $\Phi$ produces a guiding center phase space Lagrangian with a simple time-independent ``symplectic part." The symplectic part of $\ell^\infty_s$, $\ell_s^{\text{symp}}$, is 
\begin{align}
\ell_s^{\text{symp}}(z,\dot{z})&=\ell^\infty_s+h^\infty_s\nonumber\\
&=\left(q_s\bm{A}+m_s v_\parallel \hat{b}-\frac{m_s}{q_s}\mu \bm{W}\right)\cdot\dot{\bm{X}}\nonumber\\
&~~~~+\frac{q_s}{m_s}\mu\,\dot{\zeta},
\end{align} and achieving this time-independence was one of the principal goals of Ref.\,\onlinecite{Miyato_2009}. The variational principle associated with $L^\infty$ is very similar to the one introduced earlier for first-order MSST drift kinetics. The action $S^\infty=\int_{t_1}^{t_2}L^\infty\,dt$ is varied subject to the constraints given by
\begin{align}
\delta V_s&=\dot{\xi}_s+L_{V_s}\xi_s\\
\delta F_s&=-\frac{1}{\mathcal{J}_s}\text{div}(\mathcal{J}_s\xi_sF_s).
\end{align}

Using this variational principle for all-orders MSST drift kinetics, the all-orders system phase space Lagrangian can finally be derived using the method described in Section\,\ref{easy}. The dressed Lagrange $1$-form and the dressed Hamiltonian must each be found.

The first step toward finding the dressed Lagrange $1$-form $\tilde{\vartheta}^\infty$ is to identify the bare Lagrange $1$-form $\vartheta^\infty$. This is done by examining the endpoint contributions to free-endpoint variations of $S^\infty$. A straightforward calculation shows that the first free-endpoint variation of $S^\infty$ about a solution of all-orders MSST drift kinetics is given by
\begin{align}
\delta S^\infty=\vartheta^\infty[\xi,j\delta\bm{E}]\big|_{t_1}^{t_2},
\end{align}
where the bare Lagrange $1$-form $\vartheta^\infty$ is given by
\begin{align}
&\vartheta^\infty[\xi,\delta j\bm{E}]=\sum_s\int \ell_s^{\text{symp}}(z,\xi_s(z))\,F_s\,d\Omega_s\nonumber\\
&-\sum_{k=1}^{\infty}\sum_{j=1}^{k}\int(\delta\partial_t^{k-j}\bm{E})\,\cdot \bigg((-\partial_t)^{j-1}\frac{\delta\mathcal{K}^\infty}{\delta\partial_t^k\bm{E}}\bigg)\,d^3\bm{X},
\end{align}
with $\mathcal{K}^\infty(F,j\bm{E})=\sum_s\int K_s^\infty(j\bm{E})\,F_s\,d\Omega_s$. In deriving this expression, the all-orders Euler-Lagrange equations,
\begin{align}
V_s^{\infty i}=&\{z^i,h_s^\infty\}_s^{\text{gc}}\label{all_orders_hamilton}\\
\nabla\cdot\bm{D}&=\rho(F)\label{all_orders_poisson}\\
\bm{D}=\epsilon_o\bm{E}-\frac{\delta\mathcal{K}^\infty}{\delta\bm{E}}&-\sum_{k=1}^{\infty}(-\partial_t)^k\frac{\delta\mathcal{K}^\infty}{\delta \partial_t^k\bm{E}},
\end{align}
must be used to eliminate all contributions to $\delta S^\infty$ except for those that come from the endpoints. The dressed Lagrange $1$-form $\tilde{\vartheta}$ can now be obtained by expressing $\vartheta$ in terms of $F$ and $\xi$ using the all-orders MSST drift kinetic system. All of the time derivatives of $\bm{E}$ must be expressed in terms of $F$ using the relationship
\begin{align}\label{no_dotEs}
j\bm{E}=j\bm{\mathcal{E}}(\mathcal{F}(F))\equiv j\tilde{\bm{\mathcal{E}}}(F).
\end{align}
Similarly, the variations of the time derivatives of $\bm{E}$ must be expressed in terms of $\xi$ using
\begin{align}
\partial_t^{k-j}\bm{E}=\bm{\mathcal{E}}^{(k-j)}(\mathcal{F}(F))\equiv\tilde{\bm{\mathcal{E}}}^{(k-j)}(F),
\end{align}
which gives
\begin{align}
\delta\partial_t^{k-j}\bm{E}&=-D\tilde{\bm{\mathcal{E}}}^{(k-j)}[\mathcal{J}^{-1}\text{div}(\mathcal{J}F\xi)]\nonumber\\
&=-\sum_{s}D_{s}\tilde{\bm{\mathcal{E}}}^{(k-j)}[\mathcal{J}^{-1}_{s}\text{div}(\mathcal{J}_{s}F_{s}\xi_{s})].
\end{align}
Here $D_s$ denotes a partial Fr\'echet derivative with respect to the species-$s$ distribution function. When this dressing procedure is complete, the all-orders dressed Lagrange $1$-form can be expressed as
\begin{align}
\tilde{\vartheta}^\infty[\xi]=\sum_s\int\tilde{\ell}_s^{\text{symp},\infty}(z,\xi_s(z);F)\,F_s\,d\Omega_s,
\end{align}
where 
\begin{align}
&\tilde{\ell}_s^{\text{symp},\infty}(z,\dot{z};F)=\ell_s^{\text{symp}}(z,\dot{z})\nonumber\\
&+\dot{z}^i\partial_{z^i}\alpha_s^\infty(F),
\end{align}
and the function $\alpha_s^\infty(F):P\rightarrow\mathbb{R}$ is given by
\begin{align}
&\alpha_s^\infty(F)=\nonumber\\
&-\sum_{k=1}^{\infty}\sum_{j=1}^{k}\bigg( D_{s}\tilde{\bm{\mathcal{E}}}^{(k-j)\dagger}(-\partial_t)^{j-1}\frac{\delta\mathcal{K}^\infty}{\delta\partial_t^k\bm{E}}\bigg)
\end{align}
Here, $(-\partial_t)^{j-1}\frac{\delta\mathcal{K}}{\delta\partial_t^k\bm{E}}$ must be expressed in terms of $F$ by evaluating the functional derivative, applying the time derivative, and then using Eq.\,(\ref{no_dotEs}); and the adjoint operation, which is indicated by $\cdot^\dagger$, is defined by the following identity,
\begin{align}
\int \delta{\bm{E}}\cdot D_{s}\tilde{\bm{\mathcal{E}}}^{(k-j)}[\delta F_{s}]\,d^3\bm{X}&=\int D_{s}\tilde{\bm{\mathcal{E}}}^{(k-j)\dagger}[\delta\bm{E}]\,\delta F_{s}\,d\Omega_{s}.
\end{align} 

In terms of the dressed Lagrange $1$-form, the dressed Hamiltonian is
\begin{align}
\tilde{\mathcal{H}}^\infty=\tilde{\vartheta}^\infty[\tilde{V}^\infty]-\tilde{L}^\infty,
\end{align}
where $\tilde{V}^\infty$ and $\tilde{L}^\infty$ are given by dressing Eqs.\,(\ref{all_orders_hamilton}) and\,(\ref{bad_lag}) using Eq.\,(\ref{no_dotEs}). More explicitly, $\tilde{\mathcal{H}}^\infty$ is given by
\begin{align}
&\tilde{\mathcal{H}}(F)=\sum_s\int (K_s^\infty(j\tilde{\bm{\mathcal{E}}}(F))+\tilde{V}_s^{\infty i}\partial_{z^i}\alpha_s^\infty(F))\,F_s\,d\Omega_s\nonumber\\
&+\frac{1}{2}\int \epsilon_o\tilde{\bm{\mathcal{E}}}\cdot\tilde{\bm{\mathcal{E}}}\,d^3\bm{X}-\int \bigg(\sum_{k=0}^\infty(-\partial_t)^k\frac{\delta\mathcal{K}^\infty}{\delta \partial_t^k\bm{E}}\bigg)\cdot \tilde{\bm{\mathcal{E}}}\,d^3\bm{X}.
\end{align}
The all-orders system phase space Lagrangian is therefore
\begin{align}
\mathcal{L}^\infty(F,V)=\tilde{\vartheta}^\infty[V]-\tilde{\mathcal{H}}^\infty(F).
\end{align}
When varying this Lagrangian, the now-familiar constraints,
\begin{align}
\delta V_s&=\dot{\xi}_s+L_{V_s}\xi_s\\
\delta F_s&=-\frac{1}{\mathcal{J}_s}\text{div}(\mathcal{J}_s\xi_sF_s),
\end{align}
must be imposed.

\subsection{The truncated system phase space Lagrangian}
As it has been presented so far, the all-orders system phase space Lagrangian is a purely formal result. However, I will now show that it can be exploited to build an attractive alternative theory to conventional first-order MSST drift kinetics. This alternative theory, which I will refer to as renormalized first-order MSST drift kinetics, will have the following desireable features: (a) a system phase space with the correct size, (b) first-order (in magnetic non-uniformity) accuracy, and (c) a Lagrangian formulation.

The procedure I will use to derive renormalized MSST drift kinetics mirrors the one I used earlier to derive renormalized two-oscillator drift kinetics. First the all-orders system phase space Lagrangian must be expanded in powers of $\epsilon=\rho/L$, which is formally equivalent to expanding in powers of the inverse elementary charge, $|e|$ (note that $q_s=Z_s |e|$, where $Z_s$ is some integer.)  For the sake of achieving first-order accuracy, both the dressed Lagrange $1$-form and the dressed Hamiltonian will be expanded up to first order if $|e|^{-1}$. 

In order to calculate $\tilde{\vartheta}$ and $\tilde{\mathcal{H}}$ with the desired level of accuracy, it is necessary to have explicit expression for some parts of the operator $j\tilde{\bm{\mathcal{E}}}(F)$. To be precise, if $\tilde{\bm{\mathcal{E}}}$ and $\tilde{\bm{\mathcal{E}}}^{(1)}$ are expanded in powers of $\epsilon$ as
\begin{align}
\tilde{\bm{\mathcal{E}}}&=\tilde{\bm{\mathcal{E}}}_o+\tilde{\bm{\mathcal{E}}}_1+\dots\\
\tilde{\bm{\mathcal{E}}}^{(1)}&=\tilde{\bm{\mathcal{E}}}^{(1)}_o+\tilde{\bm{\mathcal{E}}}^{(1)}_1+\dots,
\end{align}
where subscripts denote order in $\epsilon$, $\tilde{\bm{\mathcal{E}}}_o$, $\tilde{\bm{\mathcal{E}}}_1$, and $\tilde{\bm{\mathcal{E}}}^{(1)}_o$ must be known. One way to calculate these quantities is to find the first few orders in the asymptotic expansion of the operator $\mathcal{F}$ and then evaluate the composition $j\bm{\mathcal{E}}\circ\mathcal{F}$. However, the most direct approach is to analyze the Poisson equation in the all-orders MSST Euler-Lagrange equations, namely Eq.\,(\ref{all_orders_poisson}). It is straightforward to show that the dominant balance of Eq.\,(\ref{all_orders_poisson}) is given by
\begin{align}\label{dom_bal_poisson}
\nabla\cdot\bigg(\epsilon_o\tilde{\bm{\epsilon}}\cdot\tilde{\bm{\mathcal{E}}}_o\bigg)=\rho(F),
\end{align}
where the dyad $\tilde{\bm{\epsilon}}$ is given by
\begin{align}
\tilde{\bm{\epsilon}}=1+\left(\sum_s\frac{c^2\mu_o m_s N_s}{B^2}\right)(1-\hat{b}\hat{b}).
\end{align}
Because $\nabla\times\tilde{\bm{\mathcal{E}}}_o=0$, this equation uniquely determines $\tilde{\bm{\mathcal{E}}}_o$. Indeed, if
\begin{align}
G_{\perp}=(\Delta+\Delta_\perp)^{-1},
\end{align}
where the negative-definite elliptic operator $\Delta+\Delta_\perp$ is given by $(\Delta+\Delta_\perp)\varphi=\nabla\cdot(\tilde{\bm{\epsilon}}\cdot\nabla\varphi)$,
Eq.\,(\ref{dom_bal_poisson}) prescribes $\tilde{\bm{\mathcal{E}}}_o$ as
\begin{align}
\tilde{\bm{\mathcal{E}}}_o(F)=\frac{1}{\epsilon_o}\nabla G_{\perp}[\rho(F)].
\end{align}
Note that because the operator $G_{\perp}$ depends on $F$ through the $N_s$ in $\Delta_\perp$, $\tilde{\bm{\mathcal{E}}}_o$ is a nonlinear functional of $F$. Similarly, a dominant balance of the time derivative of Eq.\,(\ref{dom_bal_poisson}) gives 
\begin{align}
\tilde{\bm{\mathcal{E}}}_o^{(1)}(F)=-\frac{1}{\epsilon_o}\nabla G_\perp[\nabla\cdot\bm{\mathfrak{J}}(F)],
\end{align}
where
\begin{align}
\bm{\mathfrak{J}}(F)=&\sum_s\bigg(q_s\langle\dot{\bm{X}}_{os}+\dot{\bm{X}}_{1s}\rangle_s N_s\nonumber\\
&-\frac{m_s}{B^2}\nabla\cdot(\langle\dot{\bm{X}}_{os}\rangle_s N_s)(1-\hat{b}\hat{b})\cdot\tilde{\bm{\mathcal{E}}}_o\bigg);
\end{align}
the angle brackets denote velocity-space averages, $\langle\bm{Q}\rangle_s=N_s^{-1}\int \bm{Q}\,F_s\,\mathcal{J}_s\,dv_\parallel\,d\mu\,d\zeta$; and
\begin{align}
\dot{\bm{X}}_{os}=& v_\parallel \hat{b}+\bm{u}_{Eo}\\
\dot{\bm{X}}_{1s}=& \bigg(\frac{u_{Eo}^2}{\omega_{cs}}\tau+\frac{2 v_\parallel}{\omega_{cs}}\hat{b}\times\bm{u}_{Eo}\cdot\bm{\kappa}-\frac{\bm{u}_{Eo}\cdot\nabla\times\bm{u}_{Eo}}{2\,\omega_{cs}}\bigg)\hat{b}\nonumber\\
&+\frac{\mu \,\hat{b}\times\nabla B}{q_s B}+\frac{v_\parallel^2 \hat{b}\times\bm{\kappa}}{\omega_{cs}}-\frac{\hat{b}\times \nabla u_{Eo}^2}{2\,\omega_{cs}},
\end{align}
with $\bm{u}_{Eo}=\tilde{\bm{\mathcal{E}}}_o(F)\times\hat{b} B^{-1}$, are the $O(1)$ and $O(\epsilon)$ contributions to the guiding center drift velocity, respectively.
Next, $\tilde{\bm{\mathcal{E}}}_1$ is determined by the first sub-dominant contribution to the all-orders Poisson equation, namely
\begin{align}\label{sub_dom_balance}
\nabla\cdot\bigg(\epsilon_o\tilde{\bm{\epsilon}}\cdot\tilde{\bm{\mathcal{E}}}_1\bigg)=-\nabla\cdot\bm{\pi}_1(F),
\end{align}
where
\begin{align}
\bm{\pi}_1(F)&=-\frac{\delta\mathcal{K}_1}{\delta\bm{E}}+\partial_t\frac{\delta\mathcal{K}_1}{\delta\dot{\bm{E}}}\\
=&-\sum_s\bigg(\frac{q_s N_s}{2\omega_{cs}^2}\dot{\bm{u}}_{Eo}-\frac{q_s}{2\omega_{cs}^2}\nabla\cdot(\langle\bm{X}_{os}\rangle_s N_s)\bm{u}_{Eo}\bigg)\nonumber\\
&-\frac{\delta\mathcal{K}_1}{\delta\bm{E}}(\tilde{\bm{\mathcal{E}}}_o,\tilde{\bm{\mathcal{E}}}_o^{(1)}),
\end{align}
$\mathcal{K}_1=\sum_s\int K_{1s}\,F_s\,d\Omega_s$,  and $\dot{\bm{u}}_{Eo}=\tilde{\bm{\mathcal{E}}}_o^{(1)}(F)\times\hat{b} B^{-1}$. Therefore 
\begin{align}
\tilde{\bm{\mathcal{E}}}_1(F)=-\frac{1}{\epsilon_o}\nabla G_\perp[\nabla\cdot\bm{\pi}_1(F)].
\end{align}
Because doing so will lead to considerable simplifications of the renormalized drift kinetic theory later on, it is also useful to introduce the ``second-order electric field'' $\tilde{\bm{\mathcal{E}}}_2(F)$, which is specified by the equation
\begin{gather}
\nabla\cdot\bigg(\epsilon_o\tilde{\bm{\epsilon}}\cdot\tilde{\bm{\mathcal{E}}}_2+\bm{\pi}(F)-\bm{\pi}_1(F)\bigg)=0,
\end{gather}
where
\begin{align}
\bm{\pi}(F)=&-\frac{\delta\mathcal{K}_1}{\delta\bm{E}}(\underline{\bm{\mathcal{E}}},\underline{\bm{\mathcal{E}}}^{(1)})+\frac{\delta^2\mathcal{K}_1}{\delta\bm{E}\delta\dot{\bm{E}}}[\underline{\bm{\mathcal{E}}}^{(1)}]\nonumber\\
&+\frac{\delta^2 \mathcal{K}_1}{\delta F\delta\dot{\bm{E}}}(\underline{\bm{\mathcal{E}}})[\dot{F}^*]
\end{align}
and
\begin{align}
\dot{F}^*_s&=-\mathcal{J}_s^{-1}\nabla\cdot(\mathcal{J}_s(\dot{\bm{X}}_{os}+\dot{\bm{X}}_{1s})F_s)\\
\underline{\bm{\mathcal{E}}}&=\tilde{\bm{\mathcal{E}}}_o+\tilde{\bm{\mathcal{E}}}_1\\
\underline{\bm{\mathcal{E}}}^{(1)}&=D\underline{\bm{\mathcal{E}}}[\dot{F}^*].
\end{align}
It is straightforward to verify that the difference $\bm{\pi}-\bm{\pi}_1=O(\epsilon^2)$, which implies that $\tilde{\bm{\mathcal{E}}}_2=O(\epsilon^2)$.

With these expressions for $\tilde{\bm{\mathcal{E}}}_o$, $\tilde{\bm{\mathcal{E}}}_1$, and $\tilde{\bm{\mathcal{E}}}^{(1)}_o$ in hand, the truncated system phase space Lagrangian can be found. The truncated dressed Lagrange $1$-form is given by
\begin{align}
\tilde{\vartheta}[\xi]=\sum_s\int\tilde{\ell}_s^{\text{symp}}(z,\xi_s(z);F)\,F_s\,d\Omega_s,
\end{align}
where 
\begin{align}
&\tilde{\ell}_s^{\text{symp}}(z,\dot{z};F)=\ell_s^{\text{symp}}(z,\dot{z})\nonumber\\
&+\dot{\bm{X}}\cdot\nabla\alpha_s,
\end{align}
the function $\alpha_s$ is given by
\begin{align}
\alpha_s&=-D_s\tilde{\bm{\mathcal{E}}}^\dagger_o\frac{\delta\mathcal{K}}{\delta\dot{\bm{E}}}\nonumber\\
&=D_s\tilde{\bm{\mathcal{E}}}^\dagger_o\bigg[\sum_{\bar{s}}\frac{m_s N_s}{2 B^2\omega_{cs}}\tilde{\bm{\mathcal{E}}}_o\times\hat{b}\bigg],
\end{align} 
and $\mathcal{K}=\mathcal{K}_o+\mathcal{K}_1=\sum_s\int (K_{os}+K_{1s}) F_s\,d\Omega_s$. The truncated dressed Hamiltonian is given by
\begin{align}
&\tilde{\mathcal{H}}(F)=\sum_s\int (\tilde{K}_s(F)F_s+\alpha_s^* \dot{F}_s^*)\,d\Omega_s\nonumber\\
&+\frac{1}{2}\int \epsilon_o\tilde{\bm{\mathcal{E}}}\cdot\tilde{\bm{\mathcal{E}}}\,d^3\bm{X}+\int \tilde{\bm{P}}\cdot \tilde{\bm{\mathcal{E}}}\,d^3\bm{X},
\end{align}
where $\tilde{\bm{\mathcal{E}}}=\tilde{\bm{\mathcal{E}}}_o+\tilde{\bm{\mathcal{E}}}_1+\tilde{\bm{\mathcal{E}}}_2$;
\begin{align}
\tilde{K}_s(F)&=K_{os}(\tilde{\bm{\mathcal{E}}})+K_{1s}(\underline{\bm{\mathcal{E}}},\underline{\bm{\mathcal{E}}}^{(1)});\\
\alpha_s^*&=-D_s\underline{\bm{\mathcal{E}}}^\dagger\frac{\delta\mathcal{K}_1}{\delta\dot{\bm{E}}}(\underline{\bm{\mathcal{E}}});
\end{align}
and the polarization density $\tilde{\bm{P}}$ is given by
\begin{align}
\tilde{\bm{P}}=-\frac{\delta\mathcal{K}_o}{\delta\bm{E}}(\tilde{\bm{\mathcal{E}}})+\bm{\pi}(F).
\end{align}
Note that $\nabla\cdot(\epsilon_o\tilde{\bm{\mathcal{E}}}+\tilde{\bm{P}})=\rho(F)$, which can be carefully applied to show that the functional derivative of $\tilde{\mathcal{H}}$ is given by
\begin{align}\label{func_ham}
\frac{\delta\tilde{\mathcal{H}}}{\delta F_s}=H_s+\sum_{\bar{s}}\eta^*_{s\bar{s}}[\dot{F}^*_{\bar{s}}]+D_s\tilde{\bm{\mathcal{E}}}_2^\dagger[\bm{\pi}],
\end{align}
where  $H_s=q_s \tilde{\varphi}+\tilde{K}_s$, $-\nabla\tilde{\varphi}=\tilde{\bm{\mathcal{E}}}$, and $\eta^*_{s\bar{s}}=D_s\alpha_{\bar{s}}^{*\dagger}-D_{\bar{s}}\alpha_s^*$.
The truncated system phase space Lagrangian is therefore
\begin{align}
\mathcal{L}(F,V)=\tilde{\vartheta}[V]-\tilde{\mathcal{H}}.
\end{align}
Upon applying the constrained variations
\begin{align}
\delta V_s&=\dot{\xi}_s+L_{V_s}\xi_s\\
\delta F_s&=-\frac{1}{\mathcal{J}_s}\text{div}(\mathcal{J}_s\xi_sF_s)
\end{align}
to this Lagrangian, the Euler-Lagrange equations that result are given by
\begin{align}\label{el_final}
V_s^{i}=\left\{z^i,\varepsilon_s+\frac{\delta\tilde{\mathcal{H}}}{\delta F_s}\right\}_s^{\text{gc}},
\end{align}
where
\begin{align}\label{eta_defined}
\varepsilon_s=&\sum_{\bar{s}}(D_s\alpha_{\bar{s}}^\dagger-D_{\bar{s}}\alpha_s)[\mathcal{J}_{\bar{s}}^{-1}\text{div}(\mathcal{J}_{\bar{s}} V_{\bar{s}}F_{\bar{s}})]\nonumber\\
\equiv&\sum_{\bar{s}}\eta_{s\bar{s}}[\mathcal{J}_{\bar{s}}^{-1}\text{div}(\mathcal{J}_{\bar{s}} V_{\bar{s}}F_{\bar{s}})].
\end{align}
Note that because $\varepsilon_s$ depends on $V$, Eq.\,(\ref{el_final}) is an implicit equation for $V$. In order to solve for $V$, first note that
\begin{align}
\varepsilon_s&=\sum_{\bar{s}}\eta_{s\bar{s}}[\{F_{\bar{s}},\varepsilon_{\bar{s}}\}_{\bar{s}}^{\text{gc}}+\{F_{\bar{s}},\delta\tilde{\mathcal{H}}/\delta F_{\bar{s}}\}_{\bar{s}}^{\text{gc}}]\nonumber\\
&\equiv \sum_{\bar{s}}\chi_{s\bar{s}}[\varepsilon_{\bar{s}}+\delta\tilde{\mathcal{H}}/\delta F_{\bar{s}}].
\end{align}
Therefore,
\begin{align}
\mathcal{D}[\varepsilon]=\chi[\delta\tilde{\mathcal{H}}/\delta F],
\end{align}
where the matrix of operators $\mathcal{D}$ is given by
\begin{align}
(\mathcal{D}[\varepsilon])_s=\sum_{\bar{s}}\mathcal{D}_{s\bar{s}}[\varepsilon_{\bar{s}}]=\sum_{\bar{s}}(\delta_{s\bar{s}}-\chi_{s\bar{s}})[\varepsilon_{\bar{s}}].
\end{align}
In Appendix\,\ref{solving_ham}, I argue that $\mathcal{D}$ should be invertible as a mapping from multi-species phase space functions into itself when $\epsilon$ is small. Thus,
\begin{align}
\varepsilon=\mathcal{D}^{-1}\chi[\delta\tilde{\mathcal{H}}/\delta F],
\end{align}
which, in conjuction with Eq.\,(\ref{el_final}), determines the Eulerian phase space fluid velocity $V$ explicitly,
\begin{align}\label{velocity_final}
V^i&=\bigg\{z^i,\mathcal{D}^{-1}\chi\bigg[\frac{\delta\tilde{\mathcal{H}}}{\delta F}\bigg]+\frac{\delta\tilde{\mathcal{H}}}{\delta F}\bigg\}^{\text{gc}}\nonumber\\
&=\bigg\{z^i,\mathcal{D}^{-1}\frac{\delta\tilde{\mathcal{H}}}{\delta F}\bigg\}^{\text{gc}}.
\end{align}
The time evolution equation for $F$ in renormalized MSST drift kinetics is therefore
\begin{align}\label{renormalized_MSST}
\partial_t F&=-\bigg\{F,\mathcal{D}^{-1}\frac{\delta\tilde{\mathcal{H}}}{\delta F}\bigg\}^{\text{gc}},
\end{align}
which is manifestly a first-order ODE on $F$-space. Recall that the functional derivative $\frac{\delta\tilde{\mathcal{H}}}{\delta F}$ is given in Eq.\,(\ref{func_ham}).

I will conclude this section by analyzing the linearization of renormalized MSST drift kinetics about the same type of slab equilibria used in Section\,\ref{msfail}.  Again I will assume the background magnetic field is uniform and $z$-directed. It is straightforward to show that when $F_s$ is given by Eq.\,(\ref{thermal_eq}) with $N_{os}=n(1+x/L)$, $\tilde{\bm{\mathcal{E}}}_o=\tilde{\bm{\mathcal{E}}}_1=\tilde{\bm{\mathcal{E}}}_o^{(1)}=0$, the functional derivatives $\delta\tilde{\mathcal{H}}/\delta F_s=\frac{1}{2}m_s v_\parallel^2+\mu B$, and $\chi_{s\bar{s}}[\delta\tilde{\mathcal{H}}/\delta F_{\bar{s}}]=0$. Therefore the equilibria from Section\,\ref{msfail} are indeed equilibria of renormalized MSST drift kinetics. When renormalized MSST drift kinetics is linearized about this type of equilibrium, the dynamical equations governing $z$-independent fluctuations reduce to the following pair of linear integro-differential equations,
\begin{align}
-\epsilon_o(\Delta+\Delta_\perp)\delta\dot{\varphi}_1 &=\frac{\alpha}{2}\nabla_{\perp}^2\partial_y\delta\varphi_A+\gamma\nabla_\perp^4\partial_y\delta\varphi_B\\
-\epsilon_o(\Delta+\Delta_\perp)\delta\dot{\varphi}_o &=\frac{\alpha}{2}\nabla_{\perp}^2\partial_y\delta\varphi_B,
\end{align}
where $\varphi_o$ and $\varphi_1$ are the $0^{\text{th}}$ and $1^{\text{st}}$ order contributions to the electrostatic potential, and
\begin{align}
\delta\varphi_A&=\delta\varphi_o+\delta\varphi_1\nonumber\\
&~~~+\frac{\alpha\beta}{2}\frac{G_\perp}{\epsilon_o}\partial_y^2\nabla_\perp^2\frac{G_\perp}{\epsilon_o}(\delta\varphi_1+\delta\varphi_2)\nonumber\\
&~~~-\frac{\alpha\beta}{2}\bigg(\frac{G_\perp}{\epsilon_o}\bigg)^2\nabla_\perp^2\partial_y^2(\delta\varphi_o+\delta\varphi_1-\delta\varphi_2)\\
\delta\varphi_B&=\delta\varphi_o+\delta\varphi_1-\delta\varphi_2\\
\delta\varphi_2&=-\frac{\alpha\beta}{2}\bigg(\frac{G_\perp}{\epsilon_o}\bigg)^2\nabla_\perp^2\partial_y^2\delta\varphi_o.
\end{align}
Applying the same WKB analysis to this system of equations as in Section\,\ref{msfail} shows that there are two normal modes with $k_x=0$. The two branches of the dispersion relation are given by
\begin{align}
\omega_d&=-\frac{v_*k_y}{1+\frac{x}{L}+\frac{v_{Ao}^2}{c^2}}+O(\epsilon^2)\\
\omega_s&=O(\epsilon^2).
\end{align}
The first branch, $\omega_d$, recovers the usual drift wave (with a slightly modified frequency), while the second branch represents an artificial very low frequency wave. The renormalization procedure has therefore managed to eliminate the artificial high-frequency wave, $\omega_\times$, from MSST drift kinetics and replace it with the artificial low-frequency wave, $\omega_s$. From a computational point of view, an artificial low frequency wave may be preferable to an artificial high frequency wave. For instance, in an explicit integration scheme, the time step restriction imposed by a high frequency wave would be more severe than the time step restriction imposed by a low frequency wave.

\section{Discussion}
One approach to dealing with higher-order time derivatives of the electrostatic potential that appear in the equations defining higher-order drift kinetic and gyrokinetic theories would be to first neglect the terms involving the time derivatives, solve for the potential, then use this approximate potential to evaluate the neglected time derivative. Such a manipulation, although intuitively appealing, is not obviously consistent with energy and momentum conservation. One way of interpreting the results of this paper is that they show how this ``bootstrapping" technique for dealing with higher time derivatives of the potential can be modified to be fully consistent with Lagrangian mechanics, and therefore energy and momentum conservation. 

In fact, while I have been stressing the fact that renormalized MSST drift kinetics satisfies a variational principle, renormalized MSST drift kinetics is also an infinite dimensional Hamiltonian system. The Hamiltonian functional is given by $\tilde{\mathcal{H}}$ and the field-theoretic Poisson bracket is given by
\begin{align}
&[\mathcal{F},\mathcal{G}]=\int\bigg\{\mathcal{D}^{-1}\frac{\delta\mathcal{F}}{\delta F},\mathcal{D}^{-1}\frac{\delta\mathcal{G}}{\delta F}\bigg\}^{\text{gc}}F\,d\Omega\nonumber\\
&+\int\eta\bigg[\bigg\{F,\mathcal{D}^{-1}\frac{\delta\mathcal{F}}{\delta F}\bigg\}^{\text{gc}}\bigg]\bigg\{F,\mathcal{D}^{-1}\frac{\delta\mathcal{G}}{\delta F}\bigg\}^{\text{gc}}\,d\Omega,
\end{align}
where $\mathcal{F},\mathcal{G}$ are arbitrary functions of the multi-species distribution function, and the skew-symmetric operator $\eta$ is defined in Eq.\,(\ref{eta_defined}). This Poisson bracket can be deduced directly from the phase space Lagrangian for renormalized MSST drift kinetics following the example set in Ref.\,\onlinecite{Burby_thesis_2015}, and therefore automatically satisfies the Jacobi identity. This means that, aside from eliminating unphysical rapidly varying modes, an additional benefit of employing renormalized MSST drift kinetics instead of conventional MSST drift kinetics is that the dynamically-accessible stability method\cite{Morrison_Pfirsch_1989,Burby_gvm_2015} can be applied to the renormalized theory in order to assess the stability of equilibria.

Looking ahead, two important extensions of the work presented here are (a) incorporating electromagnetic effects into the renormalization procedure and (b) identifying a collision operator with exact conservation properties for renormalized MSST drift kinetics. The second task may prove to be especially challenging because it seems likely that, due to the use of the ``bootstrapping" technique, collisional renormalized MSST drift kinetics will contain the collision operator in both the kinetic equation and the drift kinetic Poisson equation. Nevertheless, it would be interesting to know if results akin to those given in Ref.\,\onlinecite{Burby_Brizard_2015} exist for renormalized MSST drift kinetics.
\begin{acknowledgements}
This work was supported by the U. S. Department of Energy, Office of Science, Fusion Energy Sciences under  Award No. DE-FG02-86ER53223.
\end{acknowledgements}

\appendix
\section{Why does ``dressing" work?\label{dressing_why}}
At two points in the main text I applied a \emph{dressing} procedure to derive a \emph{phase space Lagrangian} for a Lagrangian system. The purpose of this appendix is to give a precise definition of the term ``phase space Lagrangian," and provide a rigorous justification for using the ``dressing" procedure to find a Lagrangian system's phase space Lagrangian. For the sake of making the discussion accessible, I will strive to convey these general ideas by way of a specific example, namely all-orders two-oscillator drift kinetics. For a similar discussion in the context of variational kinetic theories, see Ref.\,\onlinecite{Burby_thesis_2015}. For a discussion of phase space Lagrangians in the context of magnetic field line flow, see Ref.\,\onlinecite{Cary_1983}.

Suppose $\mathcal{P}$ is a linear space with coordinates $z^i$ and that $X$ is a vector field on $\mathcal{P}$. The equation $\dot{z}=X(z)$, where $z\in \mathcal{P}$, specifies an autonomous system on $\mathcal{P}$ that will be referred to as $X$. The phase space for the system $X$ is the space $\mathcal{P}$. $X$ \emph{admits a phase space Lagrangian} if there is a one-form on $\mathcal{P}$, $\vartheta=\vartheta_i \mathbf{d}z^i$, and a function on $\mathcal{P}$, $h$, such that
\begin{align}\label{admits}
\iota_X\mathbf{d}\vartheta=-\mathbf{d}h\Leftrightarrow (\vartheta_{i,j}-\vartheta_{j,i})X^j=-h_{,i}.
\end{align}
The one-form $\vartheta$ and the function $h$ are $X$'s Lagrange one-form and Hamiltonian function, respectively. Given an $X$ that admits a phase space Lagrangian, $X$'s \emph{phase space Lagrangian}, $\ell$, is the function on $\mathcal{P}\times \mathcal{P}$  given by
\begin{align}\label{gen_psl}
\ell(z,\dot{z})=\vartheta_i(z)\dot{z}^i-h(z).
\end{align} 
The justification for this terminology is the following. Given an $X$ that  admits a phase space Lagrangian, $\ell$, define the functional $S:\mathfrak{P}(P)\rightarrow\mathbb{R}$, where $\mathfrak{P}(P)$ is the space of paths, $\gamma$, through $\mathcal{P}$ parameterized by the fixed interval $[t_1,t_2]$, using the formula
\begin{align}
S(\gamma)&=\int_\gamma\vartheta -\int_{t_1}^{t_2}h(\gamma(t))\,dt\nonumber\\
&=\int_{t_1}^{t_2}\ell(\gamma(t),\dot{\gamma(t)})\,dt.
\end{align}
If $\gamma$ is an integral curve of $X$, i.e. $\dot{\gamma}(t)=X(\gamma(t))$ for each $t\in[t_1,t_2]$, then the first fixed-endpoint variation of $S$ is zero at $\gamma$. Thus, the autonomous system $X$ satisfies a variational principle, and $\ell$ serves as the corresponding Lagrangian function. The modifier ``phase space" in ``phase space Lagrangian" corresponds to the fact that the functional $S$ takes as its argument a path in phase space space; the most familiar variational principle, namely Hamilton's principle, is phrased in terms of a functional whose argument is a path in configuration space.

A natural class of systems that admit a phase space Lagrangian consists of all systems that satisfy variational principles. I will illustrate this point in the context of all-orders two-oscillator drift kinetics. In this case, $\mathcal{P}$ is $(Q,P,\varphi,\dot{\varphi})$-space, and the vector field $X$ is given by
\begin{align}
X^Q&=(1-2\epsilon g \hat{q}_{-}(Q,\varphi))P-\epsilon\omega_2^2\dot{\varphi}\\
X^P&=-(\omega_1^2+2g\varphi)\hat{q}_{-}(Q,\varphi)\\
X^\varphi&=\dot{\varphi}\\
X^{\dot{\varphi}}&=-\omega_2^2\varphi-g\hat{q}_{-}^2(Q,\varphi).
\end{align}
This system admits the following variational principle. Let the set $\mathcal{Q}$ be $(Q,P,\varphi)$-space and let $\mathfrak{P}(\mathcal{Q})$ be the corresponding space of paths, $c=(Q,P,\varphi)$, through $\mathcal{Q}$ parameterized by $[t_1,t_2]$. Define the functional $S_o:\mathfrak{P}(\mathcal{Q})\rightarrow\mathbb{R}$ using the formula
\begin{align}
&S_o(c)=\int_{t_1}^{t_2}\bar{L}(c(t),\dot{c}(t),\ddot{\varphi}(t),\dddot{\varphi}(t))\,dt
\end{align}
where $\bar{L}$ is given in Eq.\,(\ref{aotod_lag}). If the first fixed-endpoint variation of $S_o$ vanishes at $c=(Q,P,\varphi)$, then, as described in Section\,\ref{easy}, $\gamma=(Q,P,\varphi,\dot{\varphi})$ is an integral curve of $X$. Conversely, if $\gamma=(Q,P,\varphi,\dot{\varphi})$ is an integral curve of $X$, then the first fixed-endpoint variation of $S_o$ vanishes at $c=(Q,P,\varphi)$. Note that $\bar{L}$ is \emph{not} a phase space Lagrangian, i.e. it is not of the form expressed in Eq.\,(\ref{gen_psl}). Nevertheless, $X$ does admit a phase space Lagrangian. To see this, first let $F_t:\mathcal{P}\rightarrow\mathcal{P}$ be the flow map of the vector field $X$; $F_t$ is the unique one-parameter family of maps $\mathcal{P}\rightarrow\mathcal{P}$ that satisfies $F_0(z)=z$ and $\partial_t F_t(z)=X(F_t(z))$ for each $z\in\mathcal{P}$. Using $F_t$, it is possible to construct a mapping $\text{sol}:\mathcal{P}\rightarrow \mathfrak{P}(\mathcal{P})$ according to the formula
\begin{align}
\text{sol}(z)(t)=F_{t-t_1}(z).
\end{align}
Note that $\text{sol}(z)$ is the unique path through $\mathcal{P}$ that (a) is an integral curve of the vector field $X$ and (b) passes through the point $z$ at $t=t_1$. Because any path $\gamma=(Q,P,\varphi,\dot{\varphi})$ through $\mathcal{P}$ determines a corresponding path, $\pi(\gamma)=(Q,P,\varphi),$ through $\mathcal{Q}$, $\text{sol}$ and $S_o$ can be used to construct the function on phase space $\bar{S}=S_o\circ\pi\circ\text{sol}:\mathcal{P}\rightarrow\mathbb{R}$. $\bar{S}$ may be thought of as the action $S_o$ regarded as a function of initial conditions. The phase space Lagrangian for all-orders two-oscillator drift kinetics can be derived by analyzing the derivatives of $\bar{S}$ with respect to $z$ and $t_2$. ($\bar{S}$ depends on $t_2$ because the upper limit of the time integral in the definition of $S_o$ is $t_2$.)

In order to calculate the derivatives of $\bar{S}$, it will be convenient to introduce the curves $\gamma_\epsilon=\text{sol}(z+\epsilon v)=(Q_\epsilon,P_\epsilon,\varphi_\epsilon,\dot{\varphi}_\epsilon)$, $c_\epsilon=\pi(\gamma_\epsilon)$, $\Gamma_\epsilon$, and $\Upsilon_\epsilon,$ where $\epsilon$ is a small positive real number, $v\in\mathcal{P}$ is an arbitrary four-component vector, and
\begin{align}
&\Gamma_\epsilon(t)=\bigg(Q_\epsilon(t),P_\epsilon(t),\varphi_\epsilon(t),\nonumber\\
&~~~~~~~~~~~~~\frac{dQ_\epsilon}{dt},\frac{d P_\epsilon}{dt},\frac{d\varphi_\epsilon}{dt},\frac{d^2\varphi_\epsilon}{dt^2},\frac{d^3\varphi_\epsilon}{dt^3}\bigg)\\
&\Upsilon_\epsilon(t)=\bigg(Q_\epsilon(t),P_\epsilon(t),\varphi_\epsilon(t),\nonumber\\
&~~~~~~~~~~~~~\frac{dQ_\epsilon}{dt},\frac{d P_\epsilon}{dt},\frac{d^2 P_\epsilon}{dt^2},\frac{d\varphi_\epsilon}{dt},\frac{d^2\varphi_\epsilon}{dt^2},\frac{d^3\varphi_\epsilon}{dt^3}\bigg).
\end{align}
Note that because $\gamma_\epsilon$ is an integral curve of $X$ for each $\epsilon$, it must be true that
\begin{align}\label{proto_dress}
\frac{dQ_\epsilon}{dt}&=X^Q(\gamma_\epsilon(t))\\
\frac{d P_\epsilon}{dt}&=X^P(\gamma_\epsilon(t))\\
\frac{d^2 P}{dt^2}&=(X^Q\partial_Q X^P+X^\varphi\partial_\varphi X^P)(\gamma_\epsilon(t))\\
\frac{d\varphi_\epsilon}{dt}&=X^\varphi(\gamma_\epsilon(t))\\
\frac{d^2\varphi_\epsilon}{dt^2}&=\frac{d\dot{\varphi}_\epsilon}{dt}=X^{\dot{\varphi}}(\gamma_\epsilon(t))\\
\frac{d^3\varphi_\epsilon}{dt^3}&=(X^Q\partial_Q X^{\dot{\varphi}}+X^\varphi\partial_\varphi X^{\dot{\varphi}})(\gamma_\epsilon(t))\label{proto_dress_b}.
\end{align}

The directional derivative of $\bar{S}$ at $z$ along $v$ is given by
\begin{align}
\mathbf{d}\bar{S}_z(v)&=\frac{d}{d\epsilon}\bigg|_0\bar{S}(z+\epsilon v)\nonumber\\
&=\frac{d}{d\epsilon}\bigg|_0S_o(c_\epsilon),
\end{align}
which is equal to a certain free-endpoint variation of $S_o$ at a curve ($c=c_0$) that satisfies the Euler-Lagrange equations. Applying integration by parts several times and making use of the fact that $c_0$ satisfies the Euler-Lagrange equations gives
\begin{align}
&\mathbf{d}\bar{S}_z(v)\nonumber\\
&=\bigg[\vartheta_Q(\Upsilon_0(t))\delta Q(t)+\vartheta_\varphi(\Upsilon_0(t))\delta \varphi(t)\nonumber\\
&~~~~~~~+\vartheta_{\dot{\varphi}}(\Upsilon_0(t))\delta\dot{\varphi}(t)+\vartheta_{\ddot{\varphi}}(\Upsilon_0(t))\delta\ddot{\varphi}(t)\bigg]_{t_1}^{t_2},
\end{align}
where 
\begin{align}
\delta Q(t)&=\frac{d}{d\epsilon}\bigg|_0 Q_\epsilon(t)\\
\delta\varphi(t)&=\frac{d}{d\epsilon}\bigg|_0 \varphi_\epsilon(t)\\
\delta{\dot{\varphi}}(t)&=\frac{d}{d\epsilon}\bigg|_0\frac{d\varphi_\epsilon}{dt}\\
\delta \ddot{\varphi}(t)&=\frac{d}{d\epsilon}\bigg|_0\frac{d^2\varphi_\epsilon}{dt^2},
\end{align}
\begin{align}
\vartheta_Q(Z)&=\frac{\partial\bar{L}}{\partial \dot{Q}}\\
\vartheta_\varphi(Z)&=\frac{\partial \bar{L}}{\partial \dot{\varphi}}-\frac{\partial^2\bar{L}}{\partial Q\partial \ddot{\varphi}}\dot{Q}-\frac{\partial^2\bar{L}}{\partial \varphi\partial \ddot{\varphi}}\dot{\varphi}-\frac{\partial^2\bar{L}}{\partial \ddot{\varphi}^2}\dddot{\varphi}\nonumber\\
&~~~+\frac{\partial^2 \bar{L}}{\partial P\partial\dddot{\varphi}}\ddot{P}\\
\vartheta_{\dot{\varphi}}(Z)&=\frac{\partial \bar{L}}{\partial \ddot{\varphi}}-\frac{\partial^2\bar{L}}{\partial P\partial\dddot{\varphi}}\dot{P}\\
\vartheta_{\ddot{\varphi}}(Z)&=\frac{\partial \bar{L}}{\partial \dddot{\varphi}},
\end{align}
and all partial derivatives are evaluated at $Z=(Q,P,\varphi,\dot{Q},\dot{P},\ddot{P},\dot{\varphi},\ddot{\varphi},\dddot{\varphi}).$ The one-form on $Z$-space $\vartheta=\vartheta_Q\,\mathbf{d}Q+\vartheta_\varphi\,\mathbf{d}\varphi+\vartheta_{\dot{\varphi}}\,\mathbf{d}\dot{\varphi}+\vartheta_{\ddot{\varphi}}\,\mathbf{d}\ddot{\varphi}$ is the bare Lagrange one-form introduced in Section\,\ref{easy}. In terms of the bare Lagrange one-form, the directional derivative of $\bar{S}$ is
\begin{align}
\mathbf{d}\bar{S}_z(v)=[\vartheta_{\Upsilon_0(t)}(\delta\Upsilon_0(t))]_{t_1}^{t_2}.
\end{align}
This expression can be simplified further by noting that, according to Eqs.\,(\ref{proto_dress})-(\ref{proto_dress_b}), $\Upsilon_0(t)=\mathbf{J}(\gamma_0(t))$ and $\delta\Upsilon_0(t)=D\mathbf{J}_{\gamma_0(t)}[\delta\gamma(t)]$, where $\mathbf{J}$ maps phase space, $\mathcal{P}$, into $Z$-space according to $\mathbf{J}=(Q_{\mathbf{J}},P_{\mathbf{J}},\varphi_{\mathbf{J}},\dot{Q}_{\mathbf{J}},\dot{P}_{\mathbf{J}},\ddot{P}_{\mathbf{J}},\dot{\varphi}_{\mathbf{J}},\ddot{\varphi}_{\mathbf{J}},\dddot{\varphi}_{\mathbf{J}})$, with
\begin{align}
Q_{\mathbf{J}}(z)&=Q\\
P_{\mathbf{J}}(z)&=P\\
\varphi_{\mathbf{J}}(z)&=\varphi\\
\dot{Q}_{\mathbf{J}}(z)&=X^Q(z)\\
\dot{P}_{\mathbf{J}}(z)&=X^P(z)\\
\ddot{P}_{\mathbf{J}}(z)&=(X^Q\partial_Q X^P+X^\varphi\partial_{\varphi} X^P)(z)\\
\dot{\varphi}_{\mathbf{J}}(z)&=X^{\varphi}(z)\\
\ddot{\varphi}_{\mathbf{J}}(z)&=X^{\dot{\varphi}}(z)\\
\dddot{\varphi}_{\mathbf{J}}(z)&=(X^Q\partial_Q X^{\dot{\varphi}}+X^\varphi\partial_{\varphi}X^{\dot{\varphi}})(z).
\end{align}
Thus, 
\begin{align}
\vartheta_{\Upsilon_0(t)}(\delta\Upsilon_0(t))&=\vartheta_{\mathbf{J}(\gamma_0(t))}(D\mathbf{J}_{\gamma_0(t)}[\delta\gamma(t)])\nonumber\\
&=(\mathbf{J}^*\vartheta)_{\gamma_0(t)}(\delta\gamma(t))\nonumber\\
&\equiv \tilde{\vartheta}^\infty_{\gamma_0(t)}(\delta\gamma(t)),
\end{align}
where $\tilde{\vartheta}^\infty=\mathbf{J}^*\vartheta$ denotes the pullback of the one-form, $\vartheta$, on $Z$-space to phase space along the mapping $\mathbf{J}$, and
\begin{align}
\mathbf{d}\bar{S}_z(v)=[\tilde{\vartheta}_{\gamma_0(t)}(\delta\gamma(t))]_{t_1}^{t_2}.
\end{align}
A straightforward application of the definition of pullback\cite{Abraham_2008} shows that $\mathbf{J}^*\vartheta$ is precisely the dressed Lagrange one-form introduced in Section\,\ref{easy}. Finally, using the fact that
\begin{align}
\delta\gamma(t)&=\frac{d}{d\epsilon}\bigg|_0 F_{t-t_1}(z+\epsilon v)\nonumber\\
&=(DF_{t-t_1})_z[v],
\end{align}
and applying the definition of pullback once more, the directional derivative of $\bar{S}$ can be expressed as
\begin{align}\label{basic_symplectic}
\mathbf{d}\bar{S}_z(v)=(F_{t_2-t_1}^*\tilde{\vartheta}^\infty-\tilde{\vartheta}^\infty)_z(v),
\end{align}
i.e. $\mathbf{d}\bar{S}=F_{t_2-t_1}^*\tilde{\vartheta}^\infty-\tilde{\vartheta}^\infty$. 

In order to find the $t_2$ derivative of $\bar{S}$, first note that
\begin{align}
\bar{S}(z)&=\int_{t_1}^{t_2}\bar{L}(\Gamma_0(t))\,dt\nonumber\\
&=\int_{t_1}^{t_2}(\bm{j}^*\bar{L})(\gamma_0(t))\,dt\nonumber\\
&=\int_{t_1}^{t_2}(\bm{j}^*\bar{L})(F_{t-t_1}(z))\,dt,
\end{align}
where the mapping $\bm{j}=(Q_{\mathbf{J}},P_{\mathbf{J}},\varphi_{\mathbf{J}},\dot{Q}_{\mathbf{J}},\dot{P}_{\mathbf{J}},\dot{\varphi}_{\mathbf{J}},\ddot{\varphi}_{\mathbf{J}},\dddot{\varphi}_{\mathbf{J}}).$ Therefore,
\begin{align}\label{time_deriv_basic}
\frac{d}{d t_2}\bar{S}(z)=F_{t_2-t_1}^*(\bm{j}^*\bar{L})(z).
\end{align}

When Eq.\,(\ref{basic_symplectic}) is combined with Eq.\,(\ref{time_deriv_basic}), a straightforward application of exterior calculus identities given in Ref.\,\onlinecite{Abraham_2008} shows that $X$ admits a phase space Lagrangian. Indeed, 
\begin{align}
\mathbf{d}\frac{d\bar{S}}{d t_2}&=\frac{d}{d t_2}(F^*_{t_2-t_1}\tilde{\vartheta}^\infty-\tilde{\vartheta}^\infty)\nonumber\\
\Rightarrow F^*_{t_2-t_1}(\mathbf{d}\bm{j}^*\bar{L})&=F_{t_2-t_1}^*(L_X\tilde{\vartheta}^\infty)\nonumber\\
\Rightarrow -\mathbf{d}(\tilde{\vartheta}^\infty(X)-\bm{j}^*\bar{L})&=\iota_X\mathbf{d}\tilde{\vartheta}^\infty,
\end{align}
which is precisely the condition given in Eq.\,(\ref{admits}). $X$'s Lagrange one-form is given by $\tilde{\vartheta}^\infty$, while its Hamiltonian is given by $h=\tilde{\vartheta}^\infty(X)-\bm{j}^*\bar{L}$. Moreover, a careful analysis of the definitions of $\tilde{\vartheta}^\infty$ and $h$ shows the dressing procedure described in Section\,(\ref{easy}) correctly identifies the phase space Lagrangian for all-orders two-oscillator drift kinetics.

\section{Why is the operator $\mathcal{D}$ invertible? \label{solving_ham}}
The linear operator $\mathcal{D}$, which maps multi-species phase space functions into multi-species phase space functions, is defined by the expression
\begin{align}
\mathcal{D}_{s\bar{s}}[\varepsilon_{\bar{s}}]=(\delta_{s\bar{s}}-\chi_{s\bar{s}})[\varepsilon_{\bar{s}}],
\end{align}
where the operator $\chi_{s\bar{s}}$ is given by
\begin{align}\label{chi_def}
\chi_{s\bar{s}}[\varepsilon_{\bar{s}}]&=\eta_{s\bar{s}}[\{F_{\bar{s}},\varepsilon_{\bar{s}}\}_{\bar{s}}^{\text{gc}}]\nonumber\\
&=2 D_s\tilde{\bm{\mathcal{E}}}_o^\dagger\frac{\delta^2\mathcal{K}}{\delta\bm{E}\delta\dot{\bm{E}}}D_{\bar{s}}\tilde{\bm{\mathcal{E}}}_o[\{F_{\bar{s}},\varepsilon_{\bar{s}}\}_{\bar{s}}^{\text{gc}}]\nonumber\\
&+\bigg(D_s\tilde{\bm{\mathcal{E}}}_o^\dagger\frac{\delta^2\mathcal{K}}{\delta F_{\bar{s}}\delta\dot{\bm{E}}}-\frac{\delta^2\mathcal{K}}{\delta F_s\delta\dot{\bm{E}}}^\dagger D_{\bar{s}}\tilde{\bm{\mathcal{E}}}_o\bigg)[\{F_{\bar{s}},\varepsilon_{\bar{s}}\}_{\bar{s}}^{\text{gc}}].
\end{align}
Here, 
\begin{align}
&D_{\bar{s}}\tilde{\bm{\mathcal{E}}}_o[\delta F_{\bar{s}}]=\frac{1}{\epsilon_o}\nabla G_\perp[q_{\bar{s}}\delta N_{\bar{s}}]\nonumber\\
&-\frac{1}{\epsilon_o}\nabla G_\perp\bigg[\nabla\cdot\bigg(\frac{c^2\mu_o m_{\bar{s}}\delta N_{\bar{s}}}{B^2}\nabla_\perp G_\perp[\rho(F)]\bigg)\bigg]\\
&D_s\tilde{\bm{\mathcal{E}}}_o^\dagger[\bm{u}]=-\frac{q_s}{\epsilon_o}G_\perp[\nabla\cdot\bm{u}]\nonumber\\
&-\frac{c^2\mu_o m_s}{B^2}\tilde{\bm{\mathcal{E}}}_o\cdot\nabla_\perp G_\perp[\nabla\cdot\bm{u}]\\
&\frac{\delta^2\mathcal{K}}{\delta\bm{E}\delta\dot{\bm{E}}}[\delta\bm{E}]=\sum_s\frac{m_s N_s}{2 B^2\omega_{cs}} \hat{b}\times \delta\bm{E}\\
&\frac{\delta^2\mathcal{K}}{\delta F_{\bar{s}}\delta\dot{\bm{E}}}[\delta F_{\bar{s}}]=\frac{m_{\bar{s}}}{2 B^2\omega_{c\bar{s}}}\hat{b}\times\tilde{\bm{\mathcal{E}}}_o \delta N_{\bar{s}}\\
&\frac{\delta^2\mathcal{K}}{\delta F_{s}\delta\dot{\bm{E}}}^\dagger[\delta\bm{E}]=\frac{m_s}{2 B^2 \omega_{cs}}\hat{b}\times\tilde{\bm{\mathcal{E}}}_o\cdot\delta\bm{E}.
\end{align}
Consider the problem of inverting the operator $\mathcal{D}$, i.e. solving the equation
\begin{align}\label{inversion_problem}
\sum_{\bar{s}}\mathcal{D}_{s\bar{s}}[\varepsilon_{\bar{s}}]=\chi_s
\end{align}
for $\varepsilon$ given $\chi$. I will argue that this can be done provided $|e|^{-1}$ is small enough and $\rho(F)=\sum_s q_s N_s=O(1)$ (the latter condition being necessary in high-flow drift kinetics in order to guarantee $\bm{u}_E=O(1)$). 

As a first step toward solving for $\varepsilon$, I will solve for the quantity $\langle\delta\varepsilon\rangle$ as a function of $\langle\delta\chi\rangle$, where
\begin{align}
\langle\delta\varepsilon\rangle_s&=\int \{F_s,\varepsilon_s\}_s^{\text{gc}}\,\mathcal{J}_s\,dv_\parallel\,d\mu\,d\zeta\\
\langle\delta\chi\rangle_s&=\int \{F_s,\chi_s\}_s^{\text{gc}}\,\mathcal{J}_s\,dv_\parallel\,d\mu\,d\zeta.
\end{align}
Note that, according to Eq.\,(\ref{chi_def}), (a) the quantity $\eta[\delta F]$ only depends on $\delta F$ through the velocity-space integral $\int \delta F_s\mathcal{J}_s\,dv_\parallel\,d\mu\,d\zeta$ and (b) $\eta[\delta F]$ has no velocity dependence. Therefore, there is a unique operator, $\langle\eta\rangle$, mapping functions on configuration space into itself that satisfies
\begin{align}
\eta_{s\bar{s}}[\delta F_{\bar{s}}]=\langle\eta\rangle_{s\bar{s}}\bigg[\int \delta F_{\bar{s}}\mathcal{J}_{\bar{s}}\,dv_\parallel\,d\mu\,d\zeta\bigg].
\end{align}
The equation\,(\ref{inversion_problem}) can therefore be expressed as
\begin{align}\label{temp2}
\varepsilon_s-\sum_{\bar{s}}\langle\eta\rangle_{s\bar{s}}[\langle\delta\varepsilon\rangle_{\bar{s}}]=\chi_s,
\end{align}
which implies
\begin{align}\label{temp1}
\sum_{\bar{s}}\langle\mathcal{D}\rangle_{s\bar{s}}[\langle\delta\varepsilon\rangle_{\bar{s}}]=\langle\delta\chi\rangle_s,
\end{align}
where
\begin{align}
&\langle\mathcal{D}\rangle_{s\bar{s}}[\langle\delta\varepsilon\rangle_{\bar{s}}]=\nonumber\\
&\delta_{s\bar{s}}\langle\varepsilon\rangle_{\bar{s}}-\int\{F_s,\langle\eta\rangle_{s\bar{s}}[\langle\delta\varepsilon\rangle_{\bar{s}}]\}_s^{\text{gc}}\mathcal{J}_s\,dv_\parallel\,d\mu\,d\zeta.
\end{align}
I will now argue that the operator $\langle\mathcal{D}\rangle$ is invertible as a mapping from functions on configuration space into itself. The general principal I will apply asserts that operators sufficiently close to invertible operators are also invertible. The operator $\langle\mathcal{D}\rangle$ can be decomposed as $\langle\mathcal{D}\rangle=\langle\mathcal{D}\rangle_o+\langle\mathcal{D}\rangle_1$ where $\langle\mathcal{D}\rangle_o$ is the leading-order (in $\epsilon=|e|^{-1}$) contribution to $\langle\mathcal{D}\rangle$ and $\langle\mathcal{D}\rangle_1=O(\epsilon)$. $\langle\mathcal{D}\rangle$ is invertible because $\langle\mathcal{D}\rangle_o$ is invertible and $\epsilon$-close to $\langle\mathcal{D}\rangle$. To see that $\langle\mathcal{D}\rangle_o$ is invertible, first note that
\begin{align}
&\langle\mathcal{D}\rangle_{os\bar{s}}[\langle\delta\varepsilon\rangle_{\bar{s}}]=\delta_{s\bar{s}}\langle\delta\varepsilon\rangle_{\bar{s}}\nonumber\\
&-\nabla\cdot\bigg(\frac{\hat{b}\times\nabla \langle\eta\rangle_{os\bar{s}}[\langle\delta\varepsilon\rangle_{\bar{s}}]}{q_sB}N_s\bigg),
\end{align}  
where
\begin{align}
&\langle\eta\rangle_{os\bar{s}}[\langle\delta\varepsilon\rangle_{\bar{s}}]=\nonumber\\
-2q_s&\frac{G_\perp}{\epsilon_o}\bigg[\nabla\cdot\bigg(\sum_{\sigma}\frac{m_\sigma N_\sigma}{2B^2\omega_{c\sigma}}\hat{b}\times\nabla\frac{G_\perp}{\epsilon_o}[q_{\bar{s}}\langle\delta\varepsilon\rangle_{\bar{s}}]\bigg)\bigg]\nonumber\\
&~~~~~~~~~~~~~~~\equiv q_s A[q_{\bar{s}}\langle\delta\varepsilon\rangle_{\bar{s}}],
\end{align}
and $A$ is an $O(\epsilon)$ linear operator. Now consider solving the equation,
\begin{align}
\sum_{\bar{s}}\langle\mathcal{D}\rangle_{os\bar{s}}[\langle\delta\varepsilon\rangle_{\bar{s}}]=j_s,
\end{align}
for $\langle\delta\varepsilon\rangle$ given $j$. In terms of the operator $A$, this equation can be written
\begin{align}\label{lam}
\langle\delta\varepsilon\rangle_s-\nabla\cdot\bigg(\hat{b}\times\nabla A[{\textstyle\sum_{\bar{s}}}q_s\langle\delta\varepsilon\rangle_{\bar{s}}]\frac{N_s}{B}\bigg)=j_s,
\end{align}
which implies that the quantity $\lambda=\sum_{\bar{s}}q_{\bar{s}}\langle\delta\varepsilon\rangle_{\bar{s}}$ satisfies
\begin{align}
\lambda+L[\lambda]=\sum_{\bar{s}}q_{\bar{s}}j_{\bar{s}},
\end{align}
where $L$ is the $O(\epsilon)$ operator given by
\begin{align}
L[\lambda]=-\nabla\cdot\bigg(\hat{b}\times\nabla A[\lambda]\frac{\rho(F)}{B}\bigg).
\end{align}
Because $L=O(\epsilon)$, the operator $(1+L)$ is invertible for $\epsilon$ small enough. Thus, $\lambda=(1+L)^{-1}[\sum_{\bar{s}}q_{\bar{s}}j_{\bar{s}}]$, and, according to Eq.\,(\ref{lam}),
\begin{align}
&\langle\delta\varepsilon\rangle_s=j_s\nonumber\\
&+\nabla\cdot\bigg(\hat{b}\times\nabla A\bigg[(1+L)^{-1}[{\textstyle \sum_{\sigma}q_{\sigma}j_\sigma}]\bigg]\frac{N_s}{B}\bigg),
\end{align}
which shows that $\langle \mathcal{D}\rangle_o$ is invertible for $\epsilon$ small enough, as claimed. For reasons already mentioned, it follows that $\langle \mathcal{D}\rangle$ is invertible for $\epsilon$ small enough. Therefore, by Eq.\,(\ref{temp1}), $\langle\delta\varepsilon\rangle=\langle\mathcal{D}\rangle^{-1}[\langle\delta\chi\rangle]$. When this result is substituted into Eq.\,(\ref{temp2}), an expression for $\varepsilon$ and a function of $\chi$ emerges,
\begin{align}
\varepsilon=\chi+\langle\eta\rangle\langle\mathcal{D}\rangle^{-1}[\langle\delta\chi\rangle],
\end{align}
which shows that $\mathcal{D}$ is invertible.
\bibliography{cumulative_bib_file}




\end{document}